\documentclass[11pt]{article}

\usepackage[normalem]{ulem}
\usepackage{amsmath, amssymb,amsthm,bm,bbm,fullpage}
\usepackage{mathrsfs}
\usepackage{cite}
\usepackage{colortbl}
\usepackage[all]{xy}
\usepackage{epsfig}
\usepackage{subfigure}
\usepackage{graphics}
\usepackage{multirow}
\usepackage{lineno}
\usepackage{setspace}
\usepackage{caption}

\usepackage{graphicx}

\theoremstyle{remark} 
	
	\doublespace

\begin{document}

\begin{centering}
{\huge
\textbf{Group reciprocity and the evolution of stereotyping}
}
\bigskip
\\
Alexander J. Stewart$^{1,*}$, and Nichola Raihani$^{2}$
\\
\bigskip
\end{centering}
\begin{flushleft}
{\footnotesize
$^1$ School of Mathematics and Statistics, University of St Andrews, St Andrews, UK
\\
$^2$ Department of Experimental Psychology, University College London, London, UK
\\
$^*$ E-mail: ajs50@st-andrews.ac.uk
}
\end{flushleft}

\noindent \emph{Stereotyping is an unavoidable feature of the way humans see each another. As such, understanding how stereotypes are formed, and when they turn negative, can have important consequences for tackling prejudice, discrimination and inter-group conflict. We study the cultural evolution of stereotyping through changes in the willingness of people to cooperate with one another based solely on their group identity. We show that stereotyping often leads to negative judgement bias in which people are overly pessimistic about the willingness of those they stereotype to cooperate. We also show that high cognitive load, or sudden economic crises, can make stereotyping more widespread and more negative. However under more benign conditions, stereotypes can evolve to be positive, and inter-group cooperation can flourish.}

\clearpage
\doublespace
\noindent \textbf{
 Stereotypes are generalized beliefs about groups of people, which are used to make decisions and judgments about them. Although such heuristics can be useful when decisions must be made quickly, or when information is lacking, they can also serve as the basis for prejudice and discrimination. In this paper we study the evolution of stereotypes through group reciprocity. We characterize the warmth of a stereotype as the willingness to cooperate with an individual based solely on the identity of the group they belong to. We show that when stereotypes are coarse, such group reciprocity is less likely to evolve, and stereotypes tend to be negative. We also show that, even when stereotypes are broadly positive, individuals are often overly pessimistic about the willingness of those they stereotype to cooperate. We then show that the tendency for stereotyping itself to evolve is driven by the costs of cognition, so that more people are stereotyped with greater coarseness as costs increase.  Finally we show that extrinsic ``shocks'', in which the benefits of cooperation are suddenly reduced, can cause stereotype warmth and judgement bias to turn sharply negative, consistent with the view that economic and other crises are drivers of out-group animosity.}

\section*{Introduction}

 Stereotyping, in which a set of characteristics is attributed to all members of an identity group, shapes many human social interactions \cite{Fiske93,sng_williams_neuberg_2016,NEUBERG2020245,Williams310,Fiske2002,https://doi.org/10.1348/014466608X314935,FISKE200777,doi:10.1177/09567976211045929}. Such generalizations can reflect or even exacerbate inter-group tensions, leading in the extreme to de-humanization of out-groups \cite{doi:10.1111/j.1467-9280.2006.01793.x,doi:10.1027/1864-9335/a000454,doi:10.1177/1368430211407643}.
More generally, however, stereotyping can be understood as the use of heuristics to guide social decision-making, which can often be a practical necessity \cite{Hutchison2015,SNG2020136}. If we lack information about an individual's past behavior, or if cognitive constraints are present, a combination of positive and negative stereotypes may be the only way to coordinate behavior and maintain cooperation. Indeed, both theoretical and experimental \cite{Bear936,RandNature,RandHuristic,doi:10.1073/pnas.1417904112} work have shown that, when deciding whether to cooperate, \emph{intuitive} decision-making is often preferable to careful deliberation.

Whether people use stereotypes when deciding to cooperate, or whether they take the time to learn about others as individuals, depends on a trade-off between ease of decision-making on the one hand and greater benefits from deliberation on the other \cite{Fiske93}. For stereotyping to be useful in this context, it must allow people to engage in successful cooperation, while helping them avoid losing out to free-riders and cheats \cite{doi:10.1073/pnas.1417904112}. If stereotypes are too coarse, people risk either cooperating when they should not, or withholding cooperation when it could be productive. If they abandon stereotypes altogether, they lose the ability to engage in intuitive decision-making and generate unnecessary cognitive burdens.

In this paper we study the evolution of stereotyping as a mechanism for cooperation under cognitive constraints. We consider a form of group reciprocity in which individuals make decisions about whether to cooperate with a partner based on the average observed behavior of the identity group to which the partner belongs. We explore the evolution of social circles -- i.e. the number of people who are not stereotyped, but are instead judged only by their individual behavior. We also study the evolution of stereotypes themselves -- i.e the degree of coarseness or specificity in the stereotypes people employ. 

We show that positive stereotypes, in which cooperation with members of a stereotype group is more likely than not, can be maintained if people interact with relatively few ($<100$) members of each group.
However we also show that negative judgement bias -- in which people tend to be pessimistic about the willingness of members of a stereotype group to cooperate -- is common even when stereotypes are positive.

We then show that the co-evolution of social circles and stereotype groups undergoes distinct phases, depending on the cognitive costs associated with remembering individual identities, as well as the benefits of cooperation. 
When cognitive costs are low, we show that social circles are large, and any stereotypes employed tend to be positive. When cognitive costs are intermediate, we show that social circles are smaller, stereotypes are coarser but generally positive, while judgement bias tends to become negative. Finally when costs are high, we show that social circles shrink and stereotypes become very coarse and negative. In contrast when benefits from cooperation increase, we show that stereotypes remain coarse but become more positive.

Finally we explore the impact of extrinsic shocks on attitudes to stereotypes. We show that when stereotypes are initially positive, and populations experience a ``shock'' which reduces the benefits of cooperation, stereotypes can turn negative, resulting in a loss of cooperation and an increase in negative judgement bias, with potential consequences for inter-group conflict and mass polarization \cite{Williams310}.  


\section*{Results}

In order to capture the role of stereotyping in social interactions, we assume that people may treat one another differently based on their identity/stereotype group or based on whether they are a part of a close social circle (Figure 1). When discussing the model we define the ``group'' as the set of individuals with the same stereotype who engage in social interactions with a focal individual. The ``group size'' is therefore the number of individuals from a given stereotype group who a focal player interacts with. In reality, the number of people who share a stereotype (but do not interact with a given focal individual) may be much larger than the group size of the model.
If two people belong to the same social circle, we assume that they know each other as individuals, and interact based on their direct experience of one another (direct reciprocity). In contrast, when interacting with a partner outside of their social circle, we assume that people make decisions based on stereotypes i.e. using assumptions about the identity group to which the other person belongs (group reciprocity).

We focus on cooperative social interactions taking place in a game theoretic setting, between a focal player and members of different stereotype groups. We assume that a focal player's decision to cooperate depends on their strategy, which takes account of the average behavior of the stereotype group to which their partner belongs. We capture this type of interaction through an iterated pairwise donation game \cite{Hilbe:2013aa,Stewart:2013fk,McAvoy} played in a population of total size $N$, in which $m\leq N$ players are distributed equally among $G$ stereotype groups, and the remaining $(N-m)$ players form the focal player's close social circle. We assume that the focal player interacts with their close social circle using direct reciprocity. In contrast, the focal player interacts with each of  $n=m/G$ players in a given stereotype group using group reciprocity. 
\\
\\
\noindent\textbf{Game dynamics between groups:} 
The game dynamics between stereotype groups occur through pairs of randomly drawn players deciding either to cooperate by paying a cost $C$ in order to donate a benefit $B$ to their co-player (where $C < B$), or else to defect and donate nothing. We assume that the game consists of many such interactions so that every player in the population gets the opportunity to help (i.e. cooperate with) every other member of the population, and vice versa, resulting in a total payoff for each player in each ``round'' of the game due given simply by
\\
\begin{center}
\emph{payoff from group reciprocity} $=$
\emph{total benefit received from being helped} $-$ \emph{total cost paid due to helping}
\end{center}

In addition to interactions between members of different stereotype groups, interactions may occur between members of the same social circle through direct reciprocity. And so the total payoff to an individual depends on their payoffs from group reciprocity, as well as their payoffs from direct reciprocity with members of their social circle, and on the cognitive costs of engaging in both types of interaction  \cite{Milinski:1998aa,Stevens05} (see below).

\begin{figure}
\centering
\includegraphics[width=0.75\linewidth]{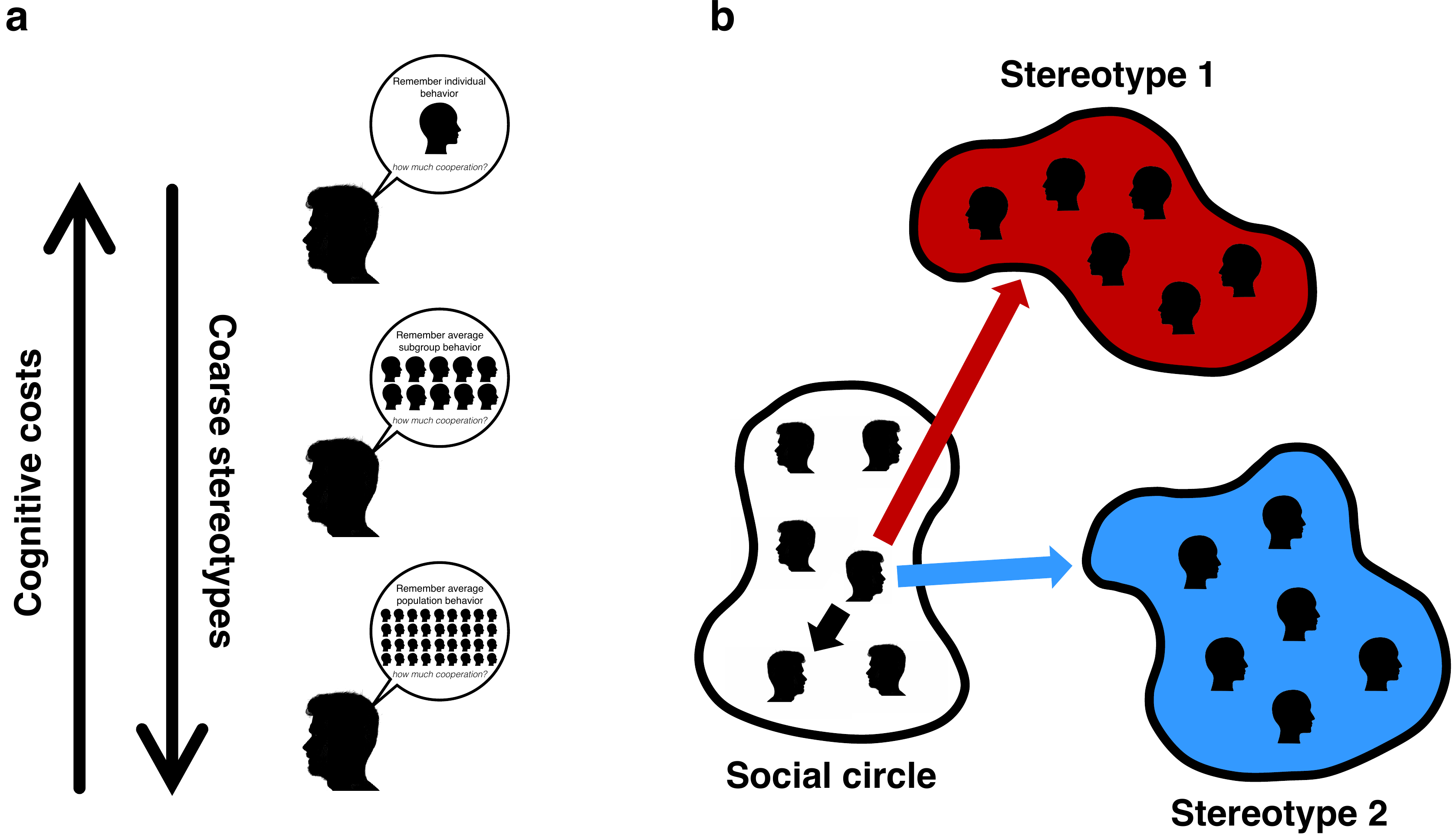}
\caption{\small \textbf{Group reciprocity and stereotyping.} a) When a player decides whether to help someone, their decision depends on how much information they have about that person --  i.e. on experience from past interactions, the ability to correctly identify other people, the ability to integrate that information to arrive at a decision and so on \cite{Doorn:2012aa}. The more information a player has about others, the better they are able to successfully employ reciprocity. In this paper we distinguish between \emph{direct} reciprocity \cite{Nowak:2006ys,Trivers}, which takes place between members of the same social circle and \emph{group} reciprocity, which takes place with members of stereotyped groups. Under direct reciprocity, players have full knowledge of each others' identity, and decide whether to cooperate based only on their direct past experience of one another. Under group reciprocity, players decide whether to cooperate based on their experience interacting with all members of the stereotype group.
Under direct reciprocity, cognitive costs are higher, but cooperation is easier to sustain, because deviations from cooperation can be dealt with more effectively. Under group reciprocity, cognitive costs are lower, but cooperation is harder to sustain, because deviations from cooperation can only be dealt with in the aggregate b) We model a population in which $m$ individuals belong to one of $G$ stereotype groups, and the rest belong to a close social circle of $(N-m)$ players. When a focal player interacts with a member of a stereotype group (red and blue interactions), they use the average behavior of that group to decide weather to cooperate (group reciprocity). When a focal player interacts with a member of their social circle (black interaction), they use the past behavior of that individual to decide weather to cooperate (direct reciprocity).}
\end{figure}

We assume that over time, players engages in a very large numbers of interaction ``rounds''. And so, in our analysis, we treat the system as an infinitely repeated donation game (see Methods).
We further assume that players can update their behavioral strategy via imitation of other players \cite{Traulsen:2006zr} (see Methods). We begin by analyzing the evolutionary dynamics of cooperation in the presence of fixed stereotype groups and in the absence of social circles ($m=N$). We then expand our analysis to consider the evolution, over longer timescales, of social circles, and finally the co-optimization of social circles and the number of stereotype groups present in the population.
\\
\\
\noindent\textbf{Stereotyping:} 
In order to study the evolution of stereotyping, we model two kinds of social interaction. First we model interactions between members of stereotype groups of size $n$, which we assume occur via \emph{group reciprocity}. Second, we also model interactions between members of the same social circle, which we assume occur in general via \emph{direct reciprocity}. We begin by studying the evolution of group reciprocity between members of fixed stereotype groups (see Methods).
We then study the evolution of stereotype groupings and social circles. Initially we assume that interactions between members of the same social are always cooperative. We relax this assumption in the SI and show that, when cooperation between members of the same circle produce lower benefits, our results are qualitatively unchanged (SI Section 3.7).

When interacting with others according to their stereotype, a focal player makes a decision to cooperate based only on their experience of that group's \emph{average} behavior. We identify the propensity of a focal individual to cooperate with a member of a group according to their stereotype of that group. Although we initially assume that this propensity is based on the experience of the focal player, we also explore scenarios in which it is derived from the average experience of \emph{all} members of the population -- which leads to a decline in the warmth of stereotypes (see SI Section 4). 

We assume that players make their decision about whether to help a given member of a given group by adopting one of a broad family of behavioral strategies, which cooperate with a probability that depends linearly on the average amount of help the player has received from members of that stereotype group in the preceding round:

\begin{equation}
p_k=s\frac{k}{n}+r
\end{equation}
\\
Here $p^i_k$ is the probability that player $i$ helps a member of of a given stereotype group of $n$ individuals, of which $k$ cooperated in the preceding round. The parameter $r$ determines the baseline rate of cooperation (i.e. the probability of cooperating even when no member of the group helped in the previous round) and $s$ determines the rate of change of cooperation with help received (i.e. the marginal increase in the probability of cooperation with each additional player who cooperated in the preceding round). In the first round we assume that players help with a probability given by their ``baseline'' rate of cooperation $r$ as given in Eq. 1, however because we are considering an infinitely repeated game with noise our analytical results are insensitive to this assumption (see Methods).

The family of conditional strategies, Eq. 1, reduces to Tit-for-Tat when $n=1$, $r=0$ and $s=1$, to always cooperate when $s=0$ and $r=1$ and to always defect when $s=r=0$. It also includes generous strategies \cite{ Nowak:2006ly,Axelrod2,Stewart:2013fk,Hilbe:2014aa}  
as well as extortionate strategies \cite{Press:2012fk, Hilbe:2013aa,Hilbe:2013uq}.

We assume that all players interact with the same number of players from a given stereotype group, $n=m/G$. We also assume that stereotyping is reciprocal, meaning that if player $i$ treats player $j$ as a stereotype, then player $j$ also treats player $i$ as a stereotype (though these players may stereotype one another in different ways). 
Eq. 1 describes a strategy for engaging in \emph{group reciprocity} between stereotype groups. We study the evolutionary dynamics of group reciprocity between a large number of such groups, with particular focus on the average rate of cooperation among groups. 

Stable cooperation requires all members of all stereotype groups adopt a strategy $s=1-r$, which simply means that a player will cooperate with certainty if everyone in the partner's stereotype group cooperated in the previous round. If such a strategy is used by all players then, when $k=n$, meaning that all players cooperated in the preceding round, then every member of each group will help every member of each other group in the next round. And so, everyone will continuously cooperate. A group in which all players use a strategy with $s=1-r$ is therefore said to be \textit{cooperative}. 

\begin{figure}
\centering
\includegraphics[width=0.45\linewidth]{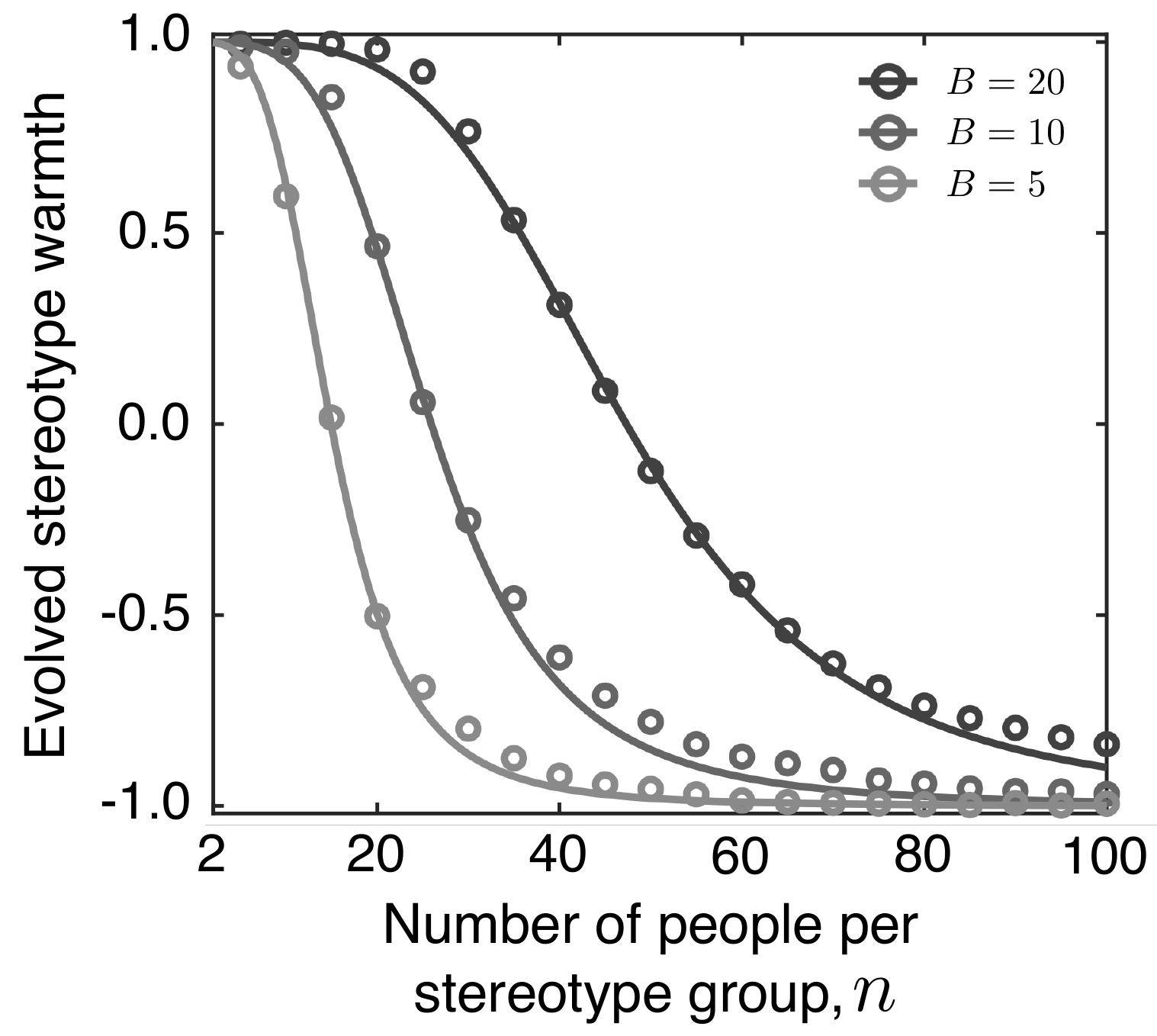}
\caption{\small \textbf{Positive and negative stereotypes.} As the number of players per-stereotype group, $n$, increases, the robustness of cooperation, $\rho$, and the resulting stereotype warmth, $W_g$ decline. We approximated the average stereotype warmth $W_g$ that will arise in a population over the course of evolution, using Eq. 3 (solid lines) for different levels of benefit from cooperation $B$, keeping the cost of cooperation fixed, $C=1$. As the ratio $B/C$ increases, the number of people per stereotype group with whom a player interacts which can sustain a positive stereotype increases. When $B/C=20$, positive stereotypes for groups of up to $\sim 50$ individuals can be sustained. When $B/C=5$, positives stereotypes for groups of only $\sim 10$ individuals can be sustained. The analytical approximation is compared to individual based simulations (dots) for stereotype groups of between 2 and 100 individuals. Results shown are for $G=10$ groups, with selection strength $\sigma=10$ and out-group imitation rate $\alpha/N=0.5$. Simulations were run for $10^4$ generations with expected cooperation rate calculated from $10^4$ sample paths. The structural constant $\beta$ was calculated numerically and has values $\beta=0.00032$ when $B=20$, $\beta=0.00073$ when $B=10$ and $\beta=0.0011$ when $B=5$ (see SI Section 5). Simulation results show the average $10^4$ replicate simulations over the course of $10^3$ generations.}
\end{figure}

Conversely, stable defection requires all players adopt a strategy $r=0$. This means that when $k=0$, no player will help any other player, and everyone will defect. A group in which all players use a strategy with $r=0$ is therefore said to be \textit{non-cooperative}.
\\
\\
\noindent \textbf{Evolution of group reciprocity:} The evolutionary dynamics among stereotype groups occurs via a process of imitation and random innovation. Players copy one another's strategy (Eq. 1) with a probability that depends on the average payoff each player received from interactions with all members of the population in the infinitely repeated game described above.  We assume that, when players update their strategy, they imitate individuals from other stereotype groups at rate $\alpha$, and otherwise imitate individuals who belong to their own stereotype group (see Methods). As a result, the probability of imitating a member of their own stereotype group is $(n-1)/((n-1)+\alpha G)$.Throughout we assume imitation of other stereotype groups occurs at rate $\alpha=0.5/N$. We explore the effects of varying $\alpha$ in the SI (Figure S4).

Under this process, the strategy space described by Eq. 1 allows for only fully cooperative, or fully non-cooperative Nash equilibria, which means only these behaviors can resist invasion \cite{Stewart:2014aa} (see Methods). Because the only available Nash equilibria are weak, the system contains no strict Evolutionary Stable Strategies, and over long time scales the system cycles between cooperation and defection (see SI Section 1) \cite{Stewart:2014aa}. Cooperative strategies, for which $s=1-r$, can resist invasion provided $s>1-\rho$ where $\rho$ describes the robustness of cooperation and is approximated by $\rho\approx \frac{1}{2}\frac{G\alpha}{n^2}\left(\frac{B}{C}-1\right)$
when $n\gg1$ (see Methods, where we also provide the full analytical form of $\rho$) -- i.e. the robustness of cooperative strategies declines rapidly with stereotype group size $n$, but increases with the rate of out-group imitation, $\alpha$, and the ratio of benefits to costs of cooperation, $B/C$. 

Similarly, non-cooperative strategies, for which $r=0$, are stable provided $s<1-\rho$. No other type of strategy can resist invasion (see Methods), and so the long term evolutionary dynamics involve repeated shifts between cooperation and defection, at a rate that depends on $\rho$ \cite{Stewart:2014aa}. Under such dynamics the long term average rate of cooperation can be approximated by \cite{Stewart:2014aa}

\begin{equation}
\Pi_c\approx\frac{\rho^2}{\rho^2+\beta(1-\rho)^2}
\end{equation}
\\
(Figure 2) where $\beta$ is a structural constant that depends on the strength of selection, $\sigma$, and can be estimated numerically \cite{Stewart:2014aa} (see Methods). We first use this approximation to study how the warmth of stereotypes change as a function of $n$, the number of people from a given stereotype group who a player interacts with. We then apply those results to study the evolution of social circle size, $(N-m)$, and the number of stereotype groups $G$.
\\

\noindent\textbf{Stereotype warmth and judgement bias:} We characterize stereotypes according to their warmth -- i.e. whether the stereotype is positive or negative about the group being considered -- and by their judgement bias -- i.e. the degree of optimism or pessimism about the group given their past actions \cite{doi:10.1098/rspb.2008.1715}. Both stereotype warmth and judgement bias are characterized in terms of the amount of cooperation between a focal player and members of a stereotype group. We define a stereotype to have a positive warmth if a player is more likely to cooperate with a member of a stereotype group than not. Specifically, we write the stereotype warmth as $W^i_g=2\Pi_c^i-1$
where $\Pi_{c}^i$ is the average rate of cooperation between a focal player $i$ and members of the group $g$. 

Over long timescales, the average stereotype warmth for the population that arises from the evolutionary dynamics described above can be approximated as $W_g\approx\frac{\rho^2-\beta(1-\rho)^2}{\rho^2+\beta(1-\rho)^2}$
(see SI Section 1).

Figure 2 shows how stereotype warmth changes with the size of the stereotype group $n$. We see that even when the benefits of cooperation are large ($B/C=20$), stereotypes become negative when players interact with more than $n\sim 50$ members of a given stereotype group. This is because group reciprocity becomes harder to maintain as stereotypes become coarser -- i.e. if a stereotype group is large, the presence of a single defector reduces the tendency of outsiders to cooperate with a large number of people --  and so lower levels of cooperation evolve at equilibrium.

In addition to stereotype warmth, we also explore the degree of judgement bias encoded in the behavioral strategies that evolve among players engaging in group reciprocity. We define a stereotype 
to have a positive judgment bias if the evolved strategy is ``optimistic'' about the behavior of members of the stereotype group. In this context we call a strategy optimistic if, for a given level of cooperation from the group $k/n$, the focal player is more likely to cooperate than they are to be cooperated with. In terms of iterated game strategies, Tit-For-Tat is neutral with respect to judgement bias since it cooperates in response to cooperation and defects in response to defection \cite{Axelrod2}. A trigger strategy has negative judgement bias, since it always defects in response to a single instance of defection \cite{8563}. A generous strategy has positive judgement bias, since it tends to cooperate even in response to defection \cite{Stewart:2012ys,Stewart:2013fk}. 

We define the judgement bias of a focal player $i$ interacting with members of a group $g$ as $J^i_g=\frac{4}{n+1}\sum_{k=0}^{n}(p_k^i-k/n)$, which in turn depends on the baseline rate of cooperation $r_i$ and the slope $s_i$ of the player's strategy (Eq. 1).
We show that the average judgement bias for the population that arises from the evolutionary dynamics at equilibrium can be approximated as $J_g\approx W_g\rho+(W_g-1)/2$
(see SI Section 2). 

\begin{figure}
\centering
\includegraphics[width=0.45\linewidth]{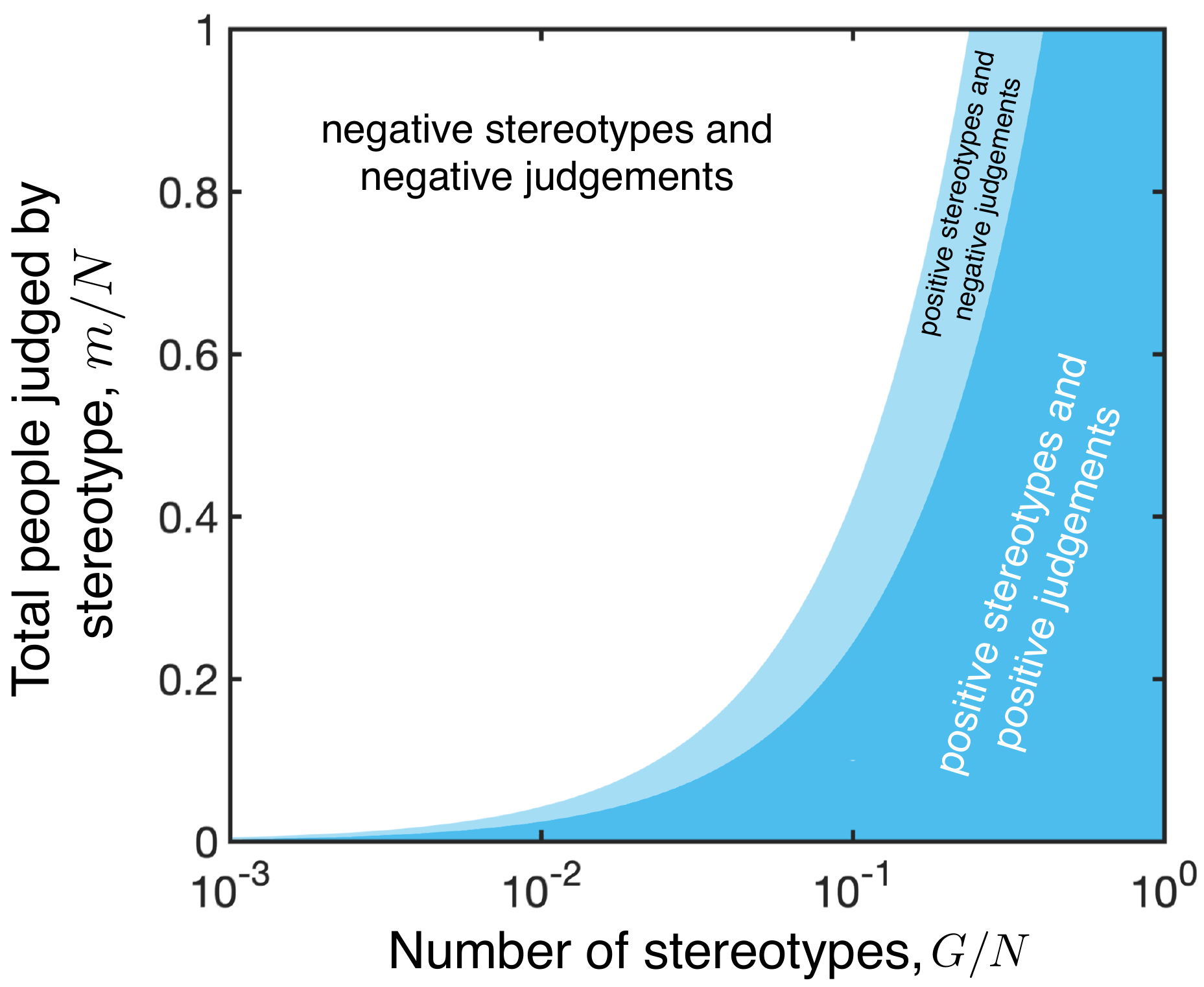}
\caption{\small \textbf{Stereotyping and judgement bias.} The evolution of positive stereotypes, $W_g>0$, and positive judgement bias, $J_g>0$ depends on the number of individuals per stereotype group, $n$. This in turn depends on the proportion of the population who are stereotyped  $m/N$ and on the number of stereotype groups $G/N$. Positive stereotype warmth (blue regions) is easier to produce than positive judgement bias (dark blue region). 
Both are easier to evolve when the number of stereotype groups is large enough
that the number of stereotyped people that a player interacts with per group is small (i.e. when the ratio $n=m/G$ is sufficiently small). This suggests when stereotyping is common, and stereotypes are coarse, attitudes towards stereotyped individuals will tend to be negative. Plots shown are based on equilibrium cooperation rates (see Methods) with $B=5$ and $C=1$.}
\end{figure}

Key to understanding this evolution is the trade-off between the efficacy of group reciprocity on the one hand (i.e. how much cooperation can be maintained among a given set of stereotype groups, as described in Figure 2-3) and the cognitive costs associated with different kinds of behavioral strategies on the other (see Figure 1).

Figure 3 shows the conditions under which positive judgement bias and positive stereotype warmth can evolve. We see that positive stereotype warmth is easier to achieve than positive judgement bias - that is, behavioral strategies that are ``optimistic'' about people are the hardest to evolve. Both positive judgment bias and stereotype warmth are easiest to evolve when the number of stereotypes, $G$, is large and the number of people being stereotyped, $m$, is small. This reflects the fact that it is only when players interact with relatively small numbers of people per stereotype group, $n=m/G$, that cooperation can be maintained (Figure 2).
\\
\\
\noindent\textbf{Cognitive capacity:} So far we have considered the evolution of group reciprocity and stereotypes for fixed social circle size $N-m$ and a fixed number of stereotype groups $G$. This assumption may be valid over timescales of a few generations, in which social attitudes may shift while the population structure remains fixed. However, over longer timescales, we must also ask how social circles and stereotype group structures themselves change.

\begin{figure}[tbhp]
\centering
\includegraphics[width=0.45\linewidth]{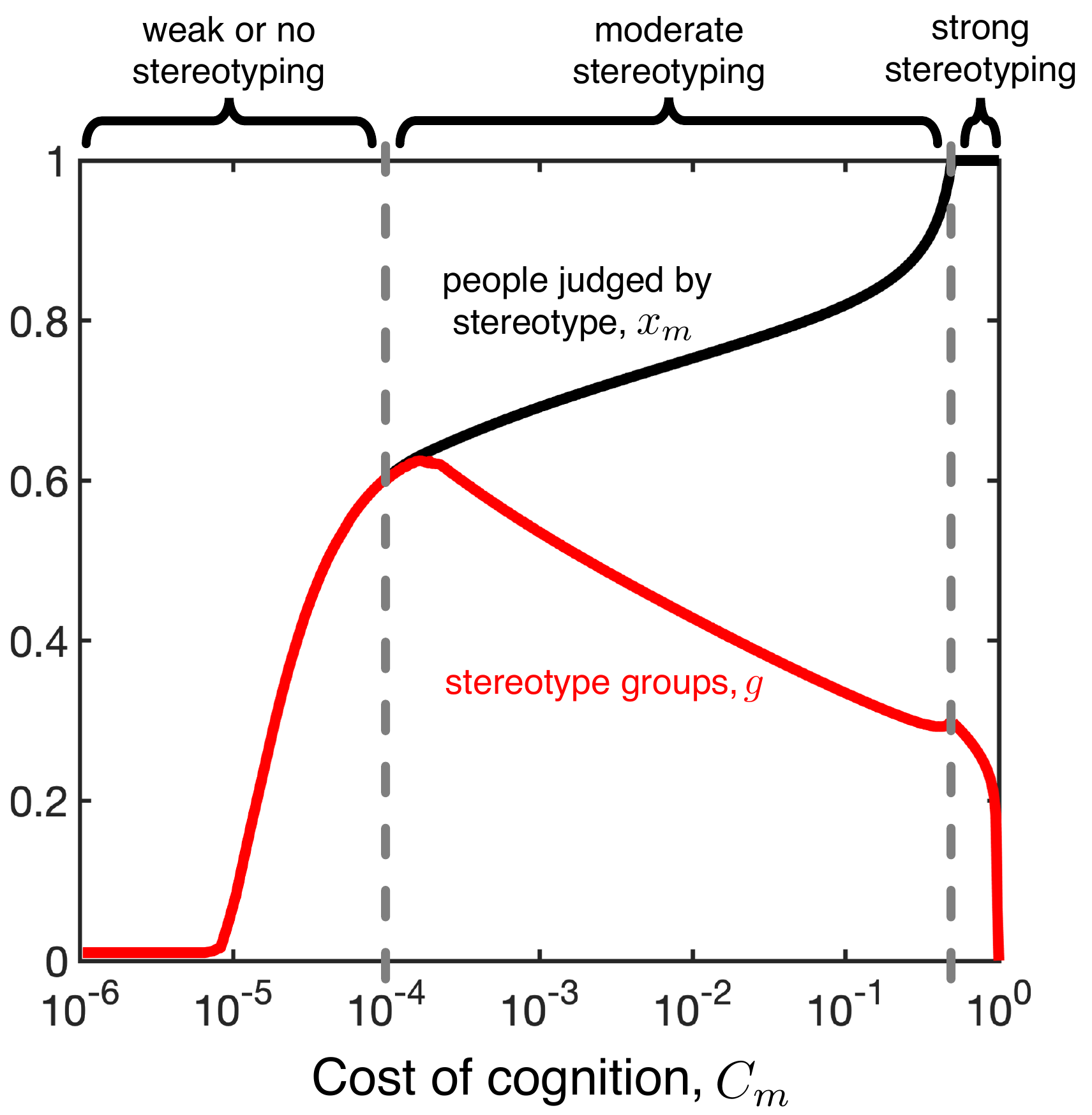}
\caption{\small \textbf{Evolutionary optimal stereotype groups and social circles.} Stereotyping takes different forms depending on the cost of cognition $C_m$ and the benefit-cost ratio of cooperation, $B/C$.  We calculated the evolutionary stable social circle size, $1-x_m$ from Eq. 4 (see Methods) as a function of the number of stereotypes, $g$, where we have set $x_m=\frac{m}{N}$ and $g=\frac{G}{N}$, and taken the limit $N\to\infty$ (see SI Section 3). We then calculated the value of $g$ that maximizes fitness, to give the evolutionary optimal social circle size and number of stereotype groups for a given set of parameters. Evolutionary optimal social circle size (black line) and number of stereotype groups (red line) as a function of cognitive costs $C_m$. When cognitive costs are small (here $C_m<10^{-4}$), there is one stereotype group per stereotyped individual, indicating weak or no stereotyping. For intermediate cognitive costs (here $0.0001<C_m<0.5$) optimal stereotypes become increasingly coarse (smaller values of $g$) and social circles shrink (higher values of $x_m$). For high cognitive costs (here $C_m>0.5$) social circles vanish ($x_m=1$) and everyone is judged via coarse stereotypes. Evolutionary optima are calculated numerically (see Methods) with $B=5$ and $C=1$, $\alpha=0.5$ and $\beta=0.0011$.} 
\end{figure}

In particular, we assume that players remember the identity and past behavior of members of their social circle, while they only remember the group identity and group average behavior of those they stereotype. The latter represents a lower cognitive cost than the former. To quantify this we calculate the information required to store the identity of each member of a social circle of size $N-m$, along with the group identities of $m$ stereotyped individuals distributed across $G$ groups i.e.

\begin{eqnarray}
\nonumber I_s(m,G)&=&\log_2 \left[A\frac{G}{N}\right]+\frac{G}{N}\log_2\left[\frac{m}{G}+1\right]\\
I_c(m)&=&\log_2 \left[A\frac{N-m}{N}\right]+\frac{N-m}{N}
\end{eqnarray}
\\
where $I_s$ is the information per population member required to store a player's group reciprocity strategy, $I_c$ is the information per population member required to store a player's strategy for interacting with their social circle. The constant $A$ scales the information required to store the identity of a given individual (see SI Section 3.1).
\\
\\
\noindent\textbf{Evolution of social circles:} In order to study the evolution of social circles we assume that we can separate the timescale of behavioral strategy evolution from the timescale of social circle evolution. In particular we assume that behavioral strategies quickly reach an equilibrium described by Eq. 2 (see SI Section 3). We then model the evolution of social circles, i.e. the proportion of players $m/N$ who are stereotyped, using the framework of adaptive dynamics \cite{Mullon:2016aa}.

Under this framework the fitness of a mutant individual $i$, who stereotypes $m_i$ individuals is given by

\begin{eqnarray}
\nonumber    w_i=&\left(1-\frac{m_i}{N}\right)(B-C)(1-C_m)^{I_{c}(m_i)}\\
    &+\frac{m_i}{N}(B-C)(1-C_m)^{I_{s}(m_i,G)}\Pi_c(n,G)
\end{eqnarray}
\\
where $C_m$ scales the cognitive cost of storing strategy and identity information about individuals and their stereotype groups,
and $\Pi_c$ is the average rate of cooperation among stereotype groups, due to the resident strategy for the population as given in Eq. 2. We have assumed that players always cooperate with members of their social circle, although we relax this assumption in the SI (Section 3.7).

In the adaptive dynamics limit $N\to\infty$ the proportion $m_i/N=x^i_m$ is a continuous variable and we can study the evolutionary dynamics of social circles by evaluating the selection gradient $\frac{\partial w_i}{\partial x^i_m}\Big |_{x^i_m=x_m}=0$ where $x_m$ is the resident value for the population.

In the supplement we show that, for a fixed number of stereotype
groups per person $G/N=g$, there is a single equilibrium social circle size (SI Section 3), with social circles tending to be smaller when cognitive costs are higher. However, we also find that there is an \emph{optimum} number of stereotype groups which maximizes population fitness (SI Section 3). In Figure 4 we study how the equilibrium social circle size, and the optimum number of stereotype groups, co-vary as a function of cognitive costs.

We find that the nature of stereotyping changes qualitatively as the cognitive costs of behavioral strategies increase. There is a threshold value of $C_m$ below which $x_m=g$ i.e. where it is optimal to have ``stereotype groups'' comprised of only a single individual. 
In this case there is really no stereotyping, because players in effect engage in direct reciprocity with ``stereotyped'' individuals. Above this threshold, there is a range of values for $C_m$ such that $x_m/g>1$ and $x_m<1$, i.e. genuine stereotyping occurs, while social circles are also maintained. As $C_m$ increases, social circles get smaller and stereotype groups decrease in number -- i.e. groups contain more people and so stereotypes become increasingly coarse. Finally there is a value of $C_m$ above which social circles vanish, i.e. $x_m=1$, and the number of stereotype groups rapidly declines.

Notably we find that increasing the relative benefits of cooperation with stereotypes, $B/C$, has similar effects on stereotyping to increasing $C_m$ (Figure S6). This occurs because, as the benefits of cooperation increase, it is easier to maintain high levels of cooperation via group reciprocity and so, for fixed cognitive costs, it is preferable to stereotype more individuals, more coarsely.
\\
\\
\noindent\textbf{Environmental shocks:} Finally we consider how stereotypes change in response to an extrinsic shock, in which the benefits of cooperation are reduced compared to historical values. We assume that over a long time scale, a population reaches an optimum level of stereotyping given the costs and benefits of cooperation and the costs of cognition. We then analyse how stereotype warmth and judgement bias shift when the benefits of cooperation are reduced, while keeping social circle size and number of stereotype groups fixed at their previous values (Figure 5). Note that the stereotypes that emerge in response to such a shock in general differ from the stereotypes that evolve when social circles and stereotype groupings are allowed to change.

We find that such  shocks tend to produce a negative shift in both stereotype warmth and  judgement bias and, most importantly, can result in positive stereotypes becoming negative. The extent of this effect depends on the pre-shock equilibrium and is most pronounced when there is a mixture of coarse stereotyping and large social circles (Figure 5c).   

\section*{Discussion}

Stereotyping is a common feature of human decision-making and is often seen as having negative social consequences \cite{Fiske93,sng_williams_neuberg_2016,NEUBERG2020245,Williams310,Fiske2002,https://doi.org/10.1348/014466608X314935,FISKE200777,doi:10.1177/09567976211045929,doi:10.1111/j.1467-9280.2006.01793.x,doi:10.1027/1864-9335/a000454,doi:10.1177/1368430211407643}. However stereotyping can produce benefits by reducing the cognitive load of decision-making, aiding coordination or signalling trust  \cite{Fiske93,Bear936,RandNature,RandHuristic,doi:10.1073/pnas.1417904112}. Here we show that stereotyping can evolve via a process of cultural evolution as a mechanism to enable cooperation while minimizing the cognitive costs of recalling the identities and past actions of large numbers of individuals. We show that, unless cognitive costs are very large (Figure 3), cultural evolution is effective at producing \emph{positive} stereotypes, which maintain cooperation among individuals who stereotype one another. However we also find the positive stereotypes can quickly turn negative after environmental ``shocks'' i.e. following a reduction in the benefit-cost ratio of cooperative interactions (Figure 4). 

This phenomenon, in which increased adversity leads to a loss of cooperation with, and increasingly negative attitudes towards, out-groups, is consistent with empirical and theoretical accounts of inter-group conflict \cite{Funke:2016,Mian:2014,doi:10.1177/1368430211407643} and a growing body of work focused on the global trend towards mass political polarization \cite{Iyengar2012,Mason2016,stewart2021inequality,Stewarteabd4201,doi:10.1146/annurev-polisci-051117-073034,doi:10.1073/pnas.2102148118}. What our results highlight is that such an increase in negative attitudes towards out-groups is a fundamental feature of the dynamics of cultural evolution, which arises when there is a mismatch between the optimal state of the system before and after an exogenous shock. Such a mismatch may be self-correcting over long timescales, if the population is able to evolve to a new cooperative optimum, in which case negative stereotyping may be transient. However in practice there may be significant inertia preventing such optimization, when it requires e.g. widespread changes in behavior or the way shared stereotype groups are defined.

\begin{figure}[tbhp]
\centering
\includegraphics[width=0.9\linewidth]{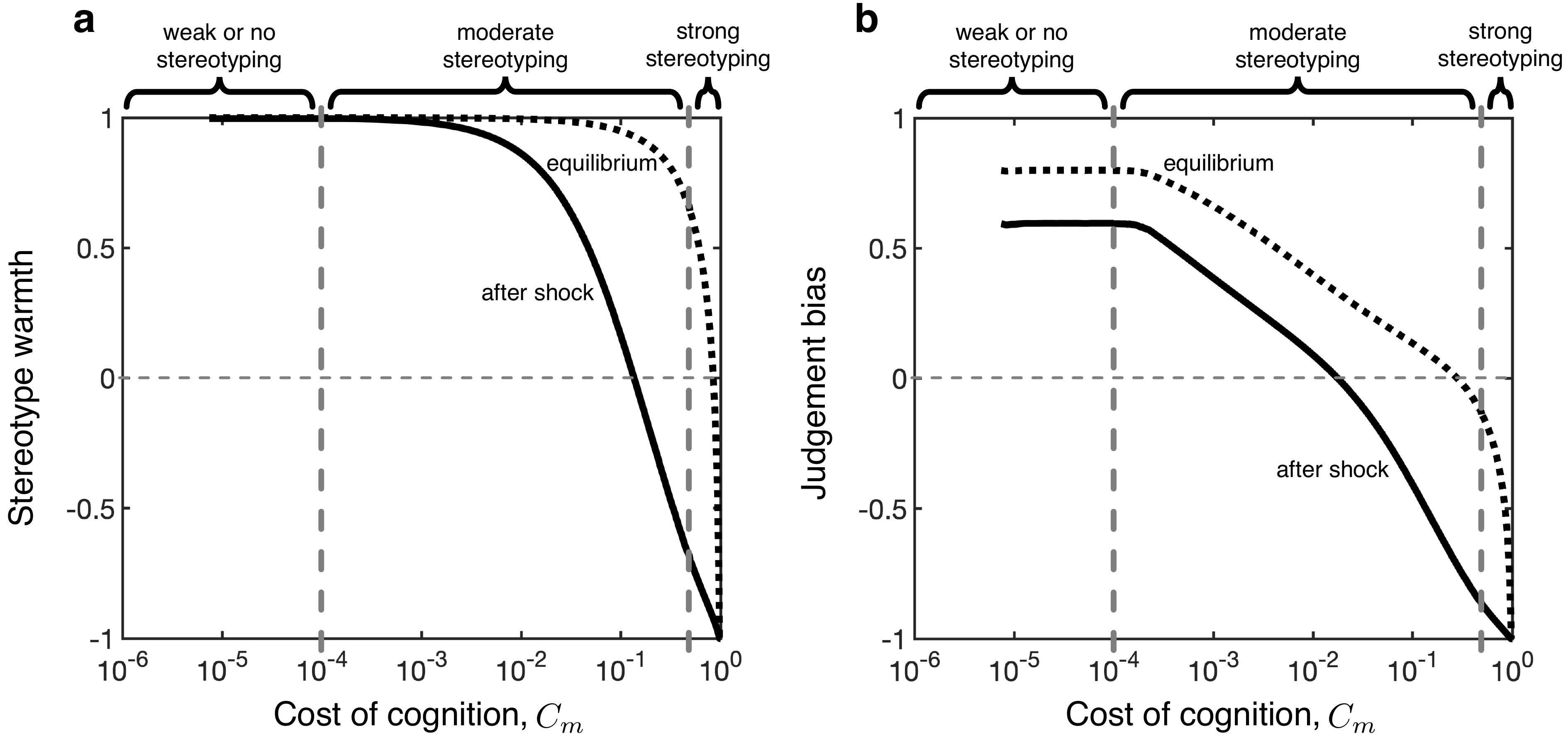}
\caption{\small \textbf{Stereotypes after environmental shocks.}  We explored what happens to stereotype warmth and judgement bias before and after an extrinsic shock, consisting of a reduction in the cost-benefit ratio from $B/C=5$ to $B/C=2.5$. We assume that social circle size, $1-x_m$, and number of stereotype groups per person, $g$, remains fixed at the evolutionary optimum for the population before the shock. We then calculate the equilibrium cooperation rate (Eq. 2) for the system before and after the shock as a function of the cost of cognition $C_m$. a) We see that stereotype warmth $W_g$ is positive before the shock (dashed line) unless the cost of cognition is very high ($C_m\sim 1$). However, after the  shock, stereotypes become more negative (solid line), and for intermediate values of $C_m$, stereotype warmth switches from being positive to negative after the shock. b) In contrast judgement bias $J_g$ becomes negative at equilibrium for low values of $C_m$ (dashed line), and becomes negative after a shock (solid line) for intermediate values of $C_m$. Evolutionary optima before and after the shock are calculated numerically with $\alpha=0.5$ and $\beta=0.0011$ (before shock) and $\beta=0.0038$ (after shock).}
\end{figure}

We interpret stereotypes through the lens of warmth -- determined by how likely an individual is to cooperate with a member of a given stereotype group -- and through judgement bias -- the degree of optimism or pessimism about the likelihood of others to cooperate based on their stereotype. Under this model, stereotype warmth reflects the realized behavior of an individual towards members of a stereotype group, while judgement bias reflects the underlying behavioral strategy of an individual when interacting with members of a stereotype group, (Eq. 1). We do not attempt to model the individual characteristics (e.g. race, religion, language) that determine membership of a given group, although we implicitly assume that such variation determines group membership. 
While willingness to cooperate and judgement bias are not identical to stereotype content, we assume that they are translated into stereotype content over time -- e.g. groups that compete for resources are less likely to cooperate and so feel less warmth towards one another \cite{FISKE200777,Fiske2002}. In this context it is notable that negative judgement bias tends to emerge before negative stereotype warmth (Figure 3 and Figure 5) meaning that high levels of cooperation can be maintained with members of a stereotype group, even when attitudes towards the group are pessimistic
 And so if judgement bias drives broad negative characterizations of members of a stereotype group, this may initially occur without loss of cooperation with members of that group.

Our work focuses on the interaction between group reciprocity and stereotypes. However a key feature of stereotyping is that it involves shared assumptions about members of a group that are disconnected from personal experience (see SI Section 4).
In the context of our model, such shared assumptions determine which stereotype group an individual is assigned to. However, we have not attempted to model baseline variation in this form of stereotype content. In particular, variation in perceived competence \cite{FISKE200777,Fiske2002} has  been shown, along with stereotype warmth, to predict stereotype content across cultural contexts \cite{https://doi.org/10.1348/014466608X314935}. While we explicitly identify the degree of cooperation with the warmth of a stereotype, we do not model variation in competence across groups. Form the perspective of our model, competence constrains the baseline willingness of individuals to engage in cooperation with different groups. And so, our results can be seen to be complementary to social psychological accounts of stereotyping. We address the evolutionary question under a simplified scenario -- when there is no variation in competence between groups, how much warmth/cooperation will evolve? Future work will naturally look to the effect of variation in competence on the evolution of group reciprocity and stereotypes.



Understanding the cause and consequence of people viewing one another as stereotypes is increasingly important, as geography ceases to limit close social interaction, different forms of identity become salient, and diverse political and social movements come into conflict. The lens of imitation dynamics and cultural evolution allows us to explore how interventions seeking to reduce inter-group conflict and negative stereotyping are likely to play out over both short and long timescales. In particular we show that, when a population is easily able to reach an evolutionary optimal state, stereotypes will often have positive warmth, and maintain high-levels of cooperation. However, if stereotype groups are inflexible, this cooperation may easily be lost in response to extrinsic shocks. And so, to prevent the negative consequences of stereotyping, it may not be necessary to discourage stereotyping altogether, but rather to encourage adaptability in the way people stereotype each other.

\section*{Methods}

Here we provide analytical results on the evolutionary robustness of cooperative and non-cooperative strategies under our model of stereotypes. Further details of simulations and the adaptive dynamics analysis can be found in the SI.
\\
\\
\textbf{Payoffs in the infinitely iterated donation game:}
We assume that players engage in an infinitely iterated, asynchronous pairwise donation game with members of a stereotype group. The first player in a given interaction chooses whether to pay a cost $C$ and donate a benefit $B$ to the second player in the interaction. We assume that this game occurs in a population such that every player has the opportunity both to donate help and to receive help from every other member of each stereotype group equally (i.e all possible pairwise interactions occur with the same probability). 

We consider a focal player $i$ who divides their partners into stereotype groups, and uses a strategy $p_k^i$ to decide whether to donate to any given member of that stereotype group. Her strategy takes account of the total number of players $k$ in the group who cooperated with her across the preceding $n=m/G$ interactions as described by Eq. 1. In any given round of interactions with the $n=m/G$ members of a stereotype group, player $i$ can choose to donate between $0$ and $mC/G$ to the group, and similarly members of the the group (from the focal player's perspective) choose to donate between $0$ and $mB/G$ to the focal player. 

And so player $i$ can treat their interactions with a partner from a given stereotype group as a two-player, infinitely iterated, $n+1$ choice game of the type studied in \cite{Stewart:2016aa} and elsewhere. If we write $v^t_{lk}$ for the probability that player $i$ donated $l$ times to members of the stereotype group and members of the stereotype group donated to the player $k$ times then the time evolution of plays in the multi-choice game is described by

\begin{equation}
v^{t+1}_{lk}=\sum_{j_p}\sum_{j_g}p^l_{j_pj_g}q^k_{j_gj_p}v^t_{j_pj_g}
\end{equation}
\\
where $p^l_{j_pj_g}$ is the probability of player $i$ making $l$ donations given that they made $j_p$ donations and members of the stereotype group made $ij_g$ donations in the preceding round, while $q^k_{j_gj_p}$ is the probability that members of the stereotype group made $k$ donations to player $i$ under the same conditions. Note that $q^k_{j_gj_p}$ in general depends on the strategies of $m/G$ different individuals and is not itself a strategy, but the effective strategy of the sub-group from the perspective of $i$. However because it is the probability of an event if we sum over all possible events (i.e. all possible donations from the stereotype group to $i$) we must have $\sum_{k=0}^{n}p^k_{j_gj_p}=1$ so that

\begin{equation}
\sum_{k}v^{t+1}_{lk}=\sum_{j_g}\sum_{j_p}p^l_{ji}v^t_{ji}
\end{equation}
\\
If we now assume a strategy $p^i_k=r+s\frac{k}{n}$ independently determines each decision to contribute (or not) on the part of $i$ over all of their $n$ interactions with the stereotype group then 

\begin{equation}
p^l_{j_gj_p}={n\choose l}\left(r+s\frac{j_g}{n}\right)^l\left(1-r-s\frac{j_g}{n}\right)^{n-l}
\end{equation}
\\
if we use Eq.8 in Eq.7, multiply both sides by $l$ and sum over $l/n$ we recover

\begin{equation}
\left<j_p\right>_{t+1}=rn+s\left<j_g\right>_t
\end{equation}
\\
where $\left<j_p\right>_{t}$ is the expected number of times the focal player contributes in round $t$ and $\left<j_g\right>_{t+1}$ is the expected number of times the group contributes. If we assume a small amount of noise in the execution of play so that the Markov chain describing the sequence of plays has a unique stationary distribution (i.e does not contain multiple absorbing states), then in an infinitely iterated game at equilibrium we have simply \cite{Press:2012fk}

\begin{equation}
\left<j_p\right>=rn+s\left<j_g\right>.
\end{equation}
\\
The expected number of donations received by $i$ at round $t$ from a given member of a stereotype group is $\left<j_g\right>_t$ and the expected number of donations made is simply $\left<j_g\right>_t$. Thus the expected average payoff to player $i$ once the game has reached equilibrium such that Eq. 10 holds is

\begin{equation}
\pi_{i}=B\left<j_g\right>-C\left<j_p\right>
\end{equation}
\\
\\
\textbf{Payoff to an invader:}
We now consider a resident strategy invading in a population comprising $G$ stereotype groups of fixed size $n$, in which all interactions between members of different groups occur via group reciprocity. In particular we consider a resident strategy

\begin{equation}
p^r_k=s_r\frac{k}{n}+r_r
\end{equation}
\\
being invaded by a mutant 

\begin{equation}
p^m_k=s_m\frac{k}{n}+r_m
\end{equation}
\\
We assume that the resident strategy is used across all stereotype groups, and ask whether the mutant can spread within the population. Under this assumption the behavior of the resident strategy within a focal stereotype group is described by

\begin{equation}
    \left<j_r\right>=r_rn+s_r\left<j_g\right>,
\end{equation}
\\
whereas the behavior of the mutant strategy is described by

\begin{equation}
    \left<j_m\right>=r_mn+s_m\left<j_g\right>.
\end{equation}
\\
Finally the behavior of other stereotype groups when interacting with the focal player is described by

\begin{equation}
    \left<j_g\right>=r_rn+s_r\frac{n-1}{n}\left<j_r\right>+s_r\frac{1}{n}\left<j_m\right>,
\end{equation}
\\
Solving Eqs. 13-15 we recover

\begin{align}
   \nonumber &\left<j_g\right>=\frac{(1 + s_r) n r_r + s_r (r_m-r_r)}{ s_r (s_r - s_m) + n - s_r^2 n}\\
    \nonumber &\left<j_r\right>=\frac{(1 + s_r) n r_r +  s_r (s_r r_m-s_m r_r )}{ s_r (s_r - s_m) + n - s_r^2 n}\\
    \nonumber &\left<j_m\right>=\frac{(1 + s_r) n (s_m r_r + r_m - s_r r_m) + s_r (s_r r_m-s_m r_r)}{ s_r (s_r - s_m) + n - s_r^2 n}\\
\end{align}
\\
The payoff received by the mutant is simply

\begin{equation}
    \pi_m=B\left<j_g\right>-C\left<j_m\right>
\end{equation}
\\
whereas the payoff received by the resident strategy within the focal stereotype group is

\begin{equation}
    \pi_r=B\left<j_g\right>-C\left<j_r\right>.
\end{equation}
\\
Finally the payoff to the resident strategy due to interactions among members of other stereotype groups when interacting with the focal stereotype groups is

\begin{equation}
    \pi^*_r=B\frac{1}{n}\left<j_m\right>+\frac{n-1}{n}\left<j_r\right>-C\left<j_r\right>
\end{equation}
and the payoff for the resident strategy when interacting with other stereotype groups is

\begin{equation}
    \pi^\dagger_r=(B-C)\frac{r_r}{1-s_r}
\end{equation}
\\
\\
\textbf{Imitation dynamics:}
We assume that cultural evolution occurs through players imitating other strategies based on payoff \cite{Traulsen:2006zr}. When a mutant is rare the observed payoff of the resident strategy among stereotype groups is approximated by

\begin{equation}
    \phi_r=\frac{n-1}{n-1+\alpha G}\pi_r+\frac{\alpha G}{n-1+\alpha G}\pi^\dagger_r
\end{equation}
\\
Where the first term describes observation of $n-1$ other members of their own stereotype group (i.e.we assume people from the same stereotype group form the basis of in-group social learning of group reciprocity) and the second term describes observation of members of other groups. In contrast the observed payoff for the mutant's own strategy is simply

\begin{equation}
    \phi_m=\pi_m
\end{equation}
\\
Under the assumed imitation dynamics a player will imitate a mutant in their own  group with probability

\begin{equation}
    f_{r\to m}=\frac{1}{1+\exp[\sigma(\phi_r-\phi_m)]}
\end{equation}
and the condition for invasion is simply $\phi_m>\phi_r$. 
\\
\\
\textbf{Evolutionary robust strategies:}
It is simple to show that only a cooperative strategy, for which $r_r+s_r=1$ or a non-cooperative strategy, for which $r_r=0$, can resist invasion. In order to see this, we first calculate $\phi_r-\phi_m$ for an arbitrary resident strategy and non-cooperative invader, $r_m=0$. Substituting from Eq. 16 we then find

\begin{equation}
   \phi_r-\phi_m=r_r\times\frac{(1 - s_m) ((1 - \gamma) s_r (B - C s_r) - C (1 - s_r^2) n)}{(1 - s_r) (n-s_r (s_m + s_r (n - 1)))} 
\end{equation}
\\
Where we have set $\gamma=\frac{n-1}{n-1+\alpha G}$. If we then calculate $\phi_r-\phi_m$ for an arbitrary resident strategy and a cooperative invader, $r_m=1-s_m$ we find

\begin{equation}
   \phi_r-\phi_m=-(1-r_r-s_r)\times\frac{(1 - s_m) ((1 - \gamma) s_r (B - C s_r) - C (1 - s_r^2) n)}{(1 - s_r) (n-s_r (s_m + s_r (n - 1)))} 
\end{equation}
\\
Eqs 24 and 25 are identical except for initial factor $r_r$ in Eq. 24 and $-(1-r_r-s_r)$ in Eq. 25. And so any strategy that is not completely cooperative or completely non-cooperative can be invaded either by a cooperative or a non-cooperative strategy.

Next we must determine the stability of fully cooperative and fully non-cooperative strategies. First we note that any pair of fully cooperative strategies always cooperate with one another, and so can replace one another via neutral drift \cite{Stewart:2014aa}. Similarly, any pair of fully non-cooperative strategies always defect against one another and can similarly replace one another via neutral drift. As a result there are no strictly Evolutionary Stable Strategies in this system, since invasions can always occur via drift. Nonetheless, fully cooperative and fully non-cooperative strategies may be \emph{evolutionary robust}, meaning that they cannot be invaded other than by neutral drift \cite{Stewart:2013fk}.

In order to determine the conditions for fully cooperative strategies to be evolutionary robust, we look at the conditions for invasion against such a resident strategy, $r_r=1-s_r$, by an arbitrary invader $r_m<1-s_m$. Substituting from Eq. 16 we find

\begin{equation}
   \phi_r-\phi_m=-(1 - s_m - r_m)\times \frac{B (1 - \gamma) s_r - C (n - s_r^2 (n-(1-\gamma)))}{n-s_r(s_m + s_r ( n-1)) }
\end{equation}
\\
and the resident strategy can resist invasion provided

\begin{equation}
    s_r>\frac{-B(1-\gamma) + \sqrt{(B(1 - \gamma))^2 + 
  4 C^2 n (n-(1 - \gamma))}}{2C (n-(1- \gamma))}
\end{equation}
\\
Similarly, for a fully non-cooperative invader, we look at the conditions for a resident strategy $r_r=0$ to resist invasion against an invader $r_m>0$. Substituting from Eq. 16 we find

\begin{equation}
     \phi_r-\phi_m= r_m\times \frac{B (1 - \gamma) s_r - C (n - s_r^2 (n-(1-\gamma)))}{n-s_r(s_m + s_r ( n-1)) }
\end{equation}
\\
and the resident strategy can resist invasion provided

\begin{equation}
    s_r<\frac{-B(1-\gamma) + \sqrt{(B(1 - \gamma))^2 + 
  4 C^2 n (n-(1 - \gamma))}}{2C (n-(1- \gamma))}
\end{equation}
\\
We can now calculate the proportion of cooperative and non-cooperative strategies that are evolutionary robust. Setting 

\begin{equation}
  \rho=1-  \frac{-B(1-\gamma) + \sqrt{(B(1 - \gamma))^2 + 
  4 C^2 n (n-(1 - \gamma))}}{2C (n-(1- \gamma))},
\end{equation}
\\
from Eq. 27 the probability that a randomly drown cooperative strategy is robust is $\rho$, while the probability that a randomly drawn non-cooperative strategy is robust is $1-\rho$. Taylor expanding Eq. 30 in $1/n$ yields the approximate expression for robustness given in the main text.
\\
\\
\textbf{Evolutionary dynamics:}
Having characterized the evolutionary robust strategies associated with the system, we can also characterise the evolutionary dynamics. In particular, under the weak mutation limit, in which new invading strategies enter a stereotype group and are either lost or go to fixation before a new invader arises, the long term evolutionary dynamics consist of long periods of quasi-stable cooperative and non-cooperative strategies \cite{Stewart:2014aa}, which are slowly eroded by drift (see Figure S2). Under these dynamics the average rate of cooperation depends on the relative robustness of cooperative and non cooperative strategies, given by $\rho$ and $1-\rho$ respectively. Under these dynamics the probability that a given individual is willing to engage in cooperation is given by Eq. 2 \cite{Stewart:2014aa}.

\clearpage

\section*{Supplementary Information}

\noindent In this supplement we present additional analysis of the evolutionary dynamics of the model, including full details of the adaptive dynamic model for the evolution of stereotypes. We also present additional numerical and simulation results.\footnote{Equation numbers in the supplement begin at 1. Where main text equations are referenced, they are referred to as such.}

\section*{Evolutionary robustness}

In order to characterize the evolutionary dynamics of stereotypes under fixed groupings (i.e fixed $n$ and fixed $m$) we adapt the results of \cite{Stewart:2014aa}. In particular, because only fully cooperative or fully non-cooperative strategies are robust to invasion under our stereotype model (see Methods) the long-run evolutionary dynamics can be approximated by a two-state Markov chain with transitions from fully cooperative ($\mathcal{C}$) to fully non-cooperative ($\mathcal{D}$) behavior and vice versa (Figure S1). In order to calculate the transition probabilities for the system we must know the rate at which fully cooperative and fully non-cooperative strategies are invaded and replaced, and the rate at which such strategies become resident.

In order to calculate the transition rates, we assume that evolution proceeds via individual fixation events, such that new strategies enter the population and either reach fixation, or are lost, before any other invader arises. Once a robust strategy resident then, under strong selection, the system evolves through neutral invasion among strategies of the same type ($\mathcal{C}$ or $\mathcal{D}$). The probability $h$ of leaving a strategy type is therefore the probability that a randomly drawn strategy replaces a randomly drawn resident of that type. For the fully cooperative strategies we have \cite{Stewart:2014aa}

\begin{figure}
\centering
\includegraphics[scale=0.3]{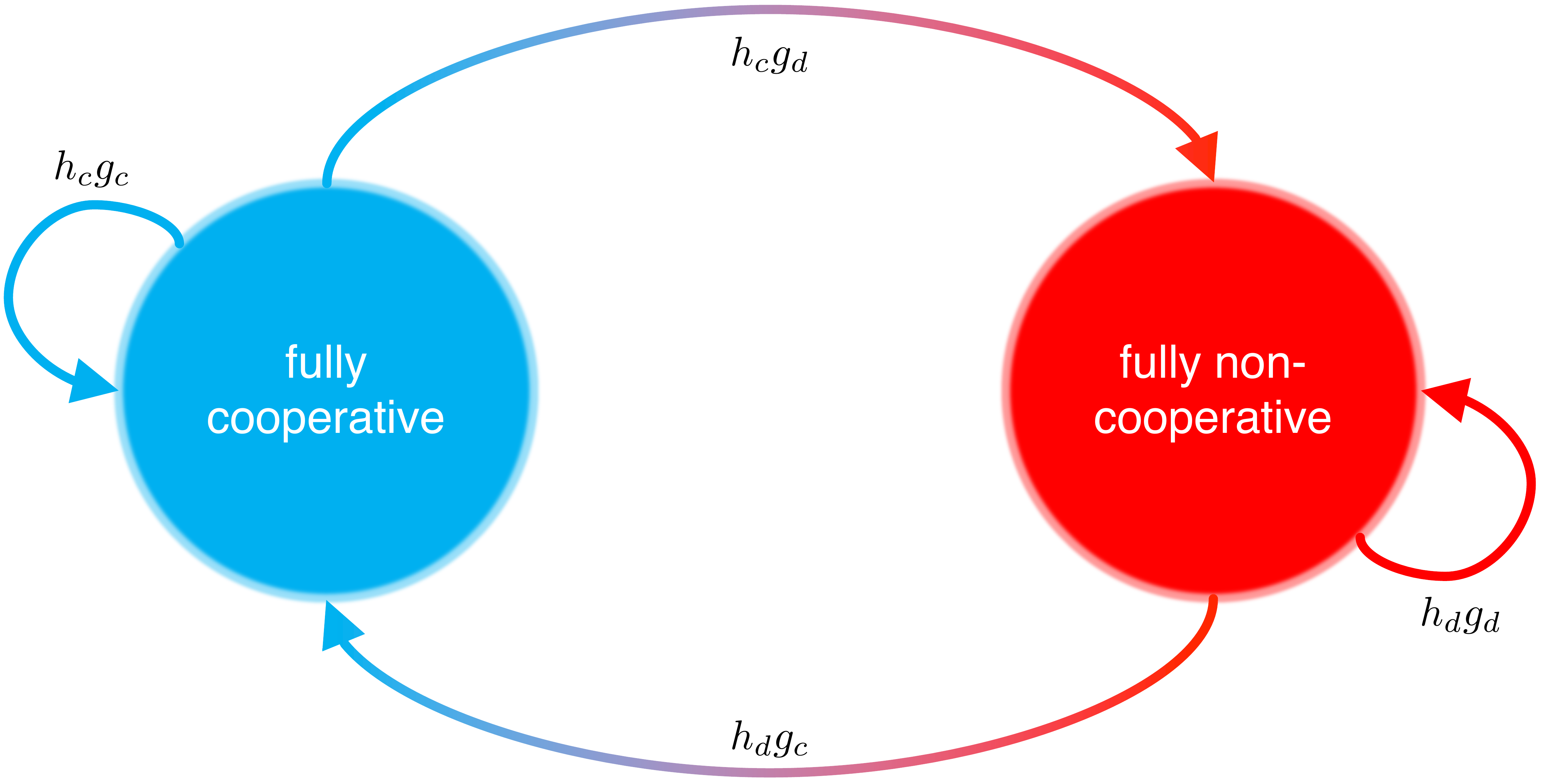}
\caption*{Figure S1: \textbf{Markov chain model.} A simplified, two-state Markov chain to describe evolution of group reciprocity. The transition rates are as given by Eqs. 1-5. In this simplified model we assume that the time spent away from these three strategy types can be neglected.}
\end{figure}

\begin{equation}
h_{c}=\int_{\mathbf{q}\in\mathcal{C}}\int_{\mathbf{p}\in\mathcal{S}}\Gamma(\mathbf{q},\mathbf{p})\mathbf{dp}\mathbf{dq}
\end{equation}
\\
where $\mathbf{q}$ is integrated over all fully cooperative strategies $\mathcal{C}$ and $\mathbf{p}$ is integrated over all possible strategies $\mathcal{S}$. The function $\Gamma(\mathbf{q},\mathbf{p})$ is the probability of fixation of a strategy $\mathbf{p}$ against resident strategy $\mathbf{q}$ under the imitation model \cite{Traulsen:2006zr}:

\begin{equation}
\Gamma(\mathbf{q},\mathbf{p})=
\left(\sum_{i=0}^{n-1}\prod_{j=1}^ie^{\sigma\left[(j-1)\phi_{pp}+(n-j)\phi_{pq}-j\phi_{qp}-(n-j-1)\phi_{qq}\right]}\right)^{-1}
\end{equation}
\\
where fixation occurs here at the level of the stereotype group of $n$ individuals.
Similarly, for the fully non-cooperative strategies we have \cite{Stewart:2014aa}

\begin{equation}
h_{d}=\int_{\mathbf{q}\in\mathcal{D}}\int_{\mathbf{p}\in\mathcal{S}}\Gamma(\mathbf{q},\mathbf{p})\mathbf{dp}\mathbf{dq}
\end{equation}
\\
When selection is strong, robust strategies cannot be invaded by strategies not of the same type, i.e $\Gamma(\mathbf{q},\mathbf{p})\to 0$ as  $\sigma\to \infty$ if $\mathbf{q}\in \mathcal{C}$ is robust and $\mathbf{p}\notin \mathcal{C}$ or if $\mathbf{q}\in \mathcal{D}$ is robust and $\mathbf{p}\notin \mathcal{D}$. And so, under the assumption that non-robust strategies are invaded quickly by one of the first alternate strategies to arise, we can approximate $h_c$ and $h_d$ as proportional to the proportion of non-robust strategies of that type, i.e. $h_c\approx (1-\rho)H_c$ and $h_d\approx \rho H_d$ where $H_c$ and $H_d$ are constant.

In order to calculate the rate at which fully cooperative and fully non-cooperative strategies become resident, we assume that the probability $g$ of entering
a strategy type is given by the probability that a robust strategy of that type
replaces a randomly drawn strategy. The probability of the population adopting a fully cooperative strategy is then \cite{Stewart:2014aa}

\begin{equation}
g_{c}=Z\delta \rho \int_{\mathbf{p}\in \mathcal{S}}\int_{\mathbf{q}\in\mathcal{C}_{r}}\Gamma(\mathbf{p},\mathbf{q})\mathbf{dp}\mathbf{dq}
\end{equation}
\\
where $\mathbf{q}$ is integrated over the set of robust fully cooperative strategies $\mathcal{C}_{r}$ and $\mathbf{p}$ is integrated over all strategies $\mathcal{S}$,
The term 
$\delta \rho$ denotes the volumes strategies within Euclidean distance $\delta$ of a fully cooperative robust strategy \cite{Sig2,Stewart:2013fk}. 
Finally $Z$ is a normalization constant so that $g_{c}+g_{d}=1$. 

The probability of the
system adopting a fully non-robust strategy is

\begin{equation}
g_{d}=Z\delta(1- \rho) \int_{\mathbf{p}\in \mathcal{S}}\int_{\mathbf{q}\in\mathcal{D}_{r}}\Gamma(\mathbf{p},\mathbf{q})\mathbf{dp}\mathbf{dq}
\end{equation}
\\
The master equation for the two-state Markov chain is then

\begin{equation}
    \Pi_c^{t+1}=(1-\Pi^t_c)h_dg_c+\Pi_c^t (1-h_c g_d)
\end{equation}
\\
where $\Pi_c^t$ is the probability that the resident strategy is fully cooperative at time $t$, where time indexes the number of invasion attempts by mutants. The stationary distribution for the system is then

\begin{equation}
    \Pi_c=\frac{h_dg_c}{h_dg_c+h_cg_d}
\end{equation}
\\
which is approximated by

\begin{equation}
    \Pi_c\approx\frac{\rho^2 H_d\int_{\mathbf{p}\in \mathcal{S}}\int_{\mathbf{q}\in\mathcal{C}_{r}}\Gamma(\mathbf{p},\mathbf{q})\mathbf{dp}\mathbf{dq}}{\rho^2 H_d \int_{\mathbf{p}\in \mathcal{S}}\int_{\mathbf{q}\in\mathcal{C}_{r}}\Gamma(\mathbf{p},\mathbf{q})\mathbf{dp}\mathbf{dq}+(1-\rho)^2 H_c\int_{\mathbf{p}\in \mathcal{S}}\int_{\mathbf{q}\in\mathcal{D}_{r}}\Gamma(\mathbf{p},\mathbf{q})\mathbf{dp}\mathbf{dq}}
\end{equation}
\\
which is main text Eq 2 with

\begin{equation}
    \beta=\frac{ H_c\int_{\mathbf{p}\in \mathcal{S}}\int_{\mathbf{q}\in\mathcal{D}_{r}}\Gamma(\mathbf{p},\mathbf{q})\mathbf{dp}\mathbf{dq}}{H_d\int_{\mathbf{p}\in \mathcal{S}}\int_{\mathbf{q}\in\mathcal{C}_{r}}\Gamma(\mathbf{p},\mathbf{q})\mathbf{dp}\mathbf{dq}}
\end{equation}
\\
Eq. S8 is not in general constant and in particular it varies with the costs and benefits of cooperation, $B$ and $C$. However we show (main text Figure 2) that it can be approximated numerically as a constant for fixed $B$ and $C$ (see below) and that such an approximation does a good job at capturing the evolutionary dynamics as group size varies. In addition, Figure S2 shows an example time series from individual based simulations, which shows that the two-state Markov chain model does indeed capture the long run evolutionary dynamics of the system.

\begin{figure}
\centering
\includegraphics[scale=0.15]{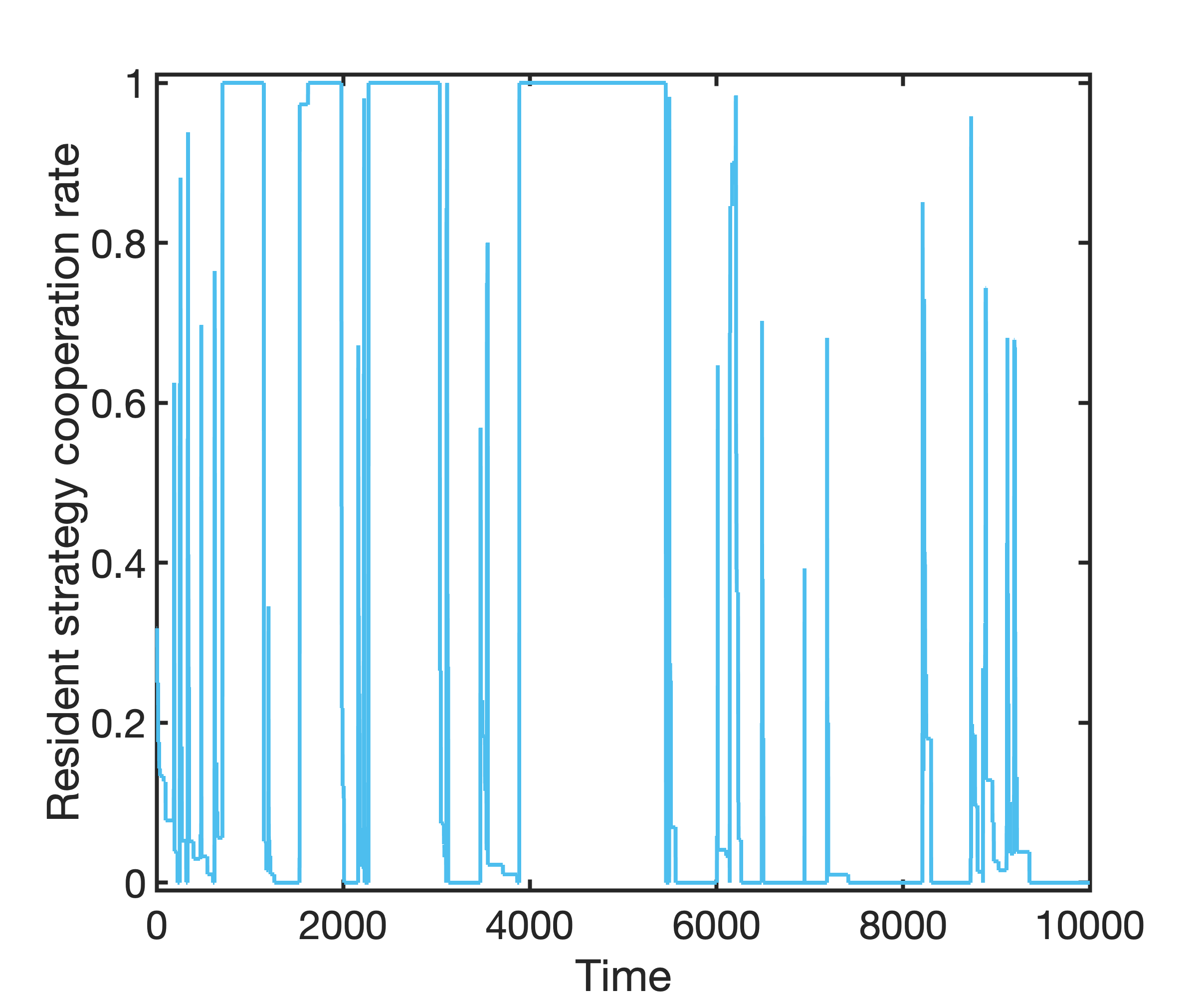}
\caption*{Figure S2: \textbf{Evolutionary dynamics time series.} Time series for a single run of the individual based simulation used to produce main text Figure 2. We see that the resident strategy moves rapidly between fully cooperative and fully non-cooperative strategies, as described in the supplementary test.}
\end{figure}
 
\section*{Stereotype warmth and Judgement bias}

We define the stereotype warmth in terms of average willingness of members of one group to cooperate with members of another group, based solely on the group identity of the latter. In particular we assume that evolutionary dynamics described above are rapid compared to the lifespan of a given individual, and so the average willingness to cooperate over time is described by $\Pi_c$. We describe a stereotype as ``warm'' if, on average, the stereotyping individual is more likely to cooperate with members of the stereotyped group than not, i.e if $\Pi_c>0.5$. We describe the stereotype as ``cool'' if, on average, the stereotyping individual is less likely to cooperate with members of the stereotyped group than not, i.e if $\Pi_c\leq 0.5$.  And so we assign a numerical value to warmth of $W_g=2\Pi_c-1$. Replacing the approximate value of $\Pi_c$ derived above (Eq.2) this becomes

\begin{equation}
    W_g=\frac{\rho^2-\beta(1-\rho)^2}{\rho^2+\beta(1-\rho)^2}
\end{equation}
\\
which is the form given in the main text and used to generate main text Figure 3. 

Judgement bias is defined as the tendency to over- or under- estimate willingness to cooperate among members of a stereotype group. In order to capture this we compare the probability of a focal individual cooperating as defined by their behavioral strategy, $p_k$, to the observed rate of cooperation among members of the stereotype group, $k/n$, for different values of $k$, i.e.

\begin{equation}
    J_g=\frac{4}{n+1}\sum_{k=0}^n\left(p_k-\frac{k}{n}\right)
\end{equation}
\\
In general for a strategy given by main text Eq. 1 this becomes

\begin{equation}
    J_g=2(s+2r-1)
\end{equation}
\\
i.e. judgement bias depends on both the slope $s$ and the baseline rate of cooperation $r$ of a player's behavioral strategy. Based on the evolutionary dynamics described above, we can estimate the average rate of judgement bias when stereotyping a group. A proportion $\Pi_c$ of the time, players use robust fully cooperative strategies, $s=1-r$ such that $s>1-\rho$. Similarly, a proportion $1-\Pi_c$ of the time, players use fully non-cooperative strategies, such that $r=0$ and $s<1-\rho$. And so the expected judgement bias is approximately

\begin{figure}
\centering
\includegraphics[scale=0.15]{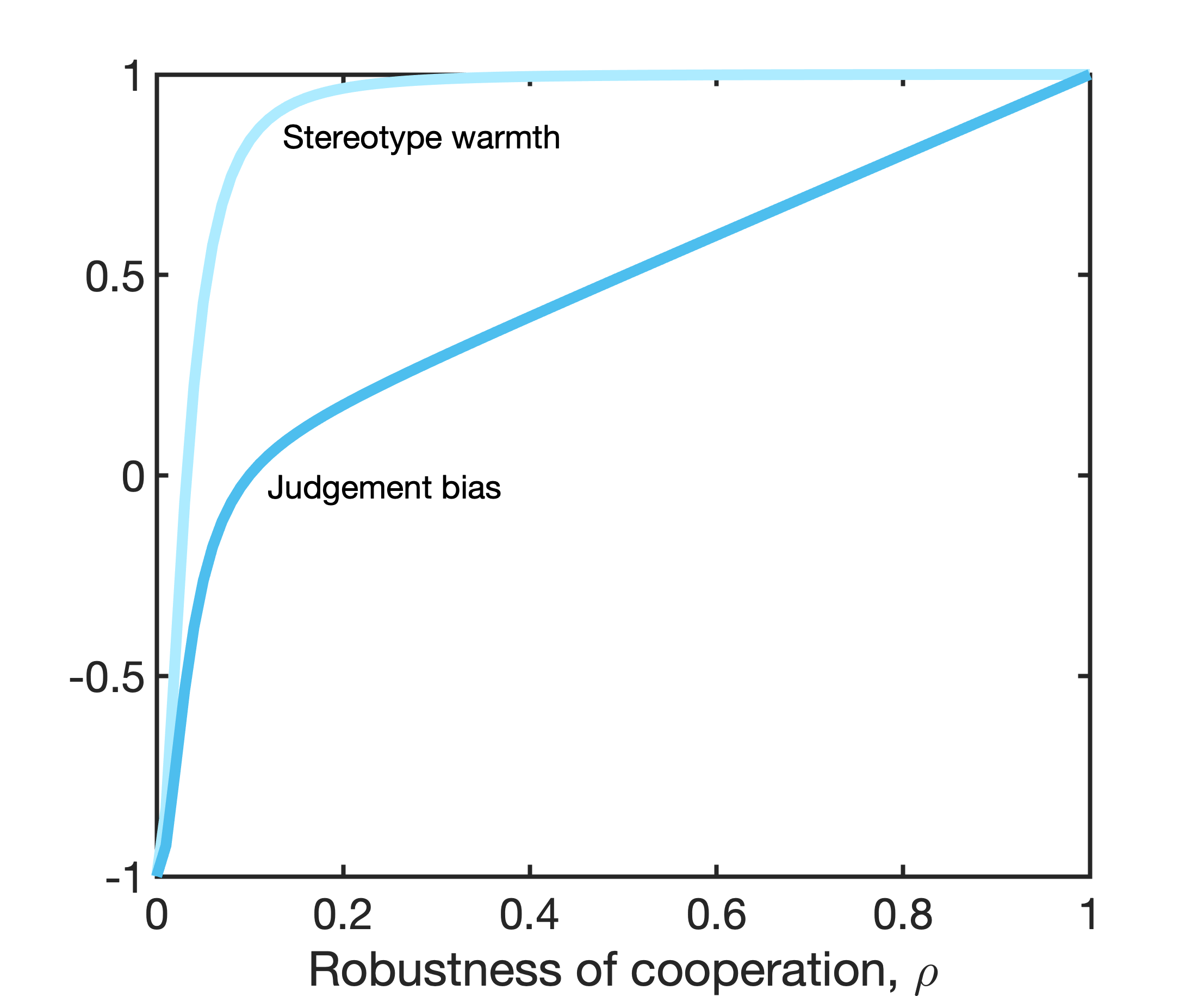}
\caption*{Figure S3 \textbf{Robustness of cooperation and stereotype content.} We calculated the stereotype warmth (darker blue) and judgement bias (lighter blue) as a function of the robustness cooperation $\rho$ from Eq. 10 and Eq. 13 with $\beta=0.0011$ corresponding to payoffs $B=5$ and $C=1$.}
\end{figure}

\begin{equation}
    \left<J_g\right>\approx\Pi_c\rho-(1-\Pi_c)(1+\rho)
\end{equation}
\\
Writing $\Pi_c=(W_g+1)/2$ we recover

\begin{equation}
    \left<J_g\right>\approx W_g\frac{1+2\rho}{2}-\frac{1}{2}
\end{equation}
\\
Note that Eq. 14 implies that when $\left<J_g\right>=0$

\begin{equation}
    W_g=\frac{1}{1+2\rho}
\end{equation}
\\
i.e. as $\rho$ decreases, average judgement bias always becomes negative before stereotype warmth (Figure S3).

\section*{The adaptive dynamics of stereotypes}

Next we discuss the adaptive dynamics of stereotype groups themselves. We assume the the evolutionary dynamics of cooperation described above occur within the lifetime of an individual via a process of social learning described by the imitation process. During this process, stereotype group size $n$ and the number of stereotype groups $G$ remain fixed and the average cooperation rate between any given individual and members of another stereotype groups is $\Pi_c$. We assume that changes to stereotype groups occur via changes in the number of stereotyped individuals $m$, and that this process occurs on a slower timescale (i.e from generation to generation).

We use the framework of adaptive dynamics to study the selection gradient acting on the number of stereotyped individuals. In order to perform this analysis we make a number of additional simplifying assumptions. First we take the limit that the total population of interacting players is very large, $N\to\infty$, in such a way that the ratio $m/N\to x_m$ and $G/N\to g$ remain finite. We then study the selection gradient acting on $x_m$. We assume that the evolutionary dynamics within an individual's lifetime continue to be described as above, so that the average rate of cooperation between groups is described by main text Eq. 2 in the limit of an infinite population, with finite stereotype groups of size $n=m/g$.

The fitness of a given individual $i$ is then given by main text Eq. 4. The first term describes the cognitive costs associated with remembering the identity of the $(1-x_m)$ people within an individual's social circle. The number of bits of information, $I_c$, required to keep track of one's social circle is given by main text Eq. 3 and the resulting cognitive cost causes an exponential decrease in fitness  with the information content of that strategy: $(1-C_m)^{I_c}$. We initially assume that within a social circle all individuals engage in successful direct reciprocity and generate payoff $(B-C)$. Similarly, the payoff associated with interactions with stereotyped individuals come at a cost $(1-C_m)^{I_s}$ but generate a benefit from group reciprocity $(B-C)\Pi_c$. Note that, for a mutant individual $i$ which changes the size of their social circle $1-x^i_m$ in a population with resident strategy $x_m$, the average rate of cooperation experienced by that individual $\Pi_c$ is unchanged, i.e. $\frac{\partial \Pi_c}{\partial x_m^i}=0$.

\subsection*{Information content of strategies}

In order to model the cost of cognition associated with different social circle sizes, we begin by considering the information required to keep track of a population of $N$ other individuals. Each player divides that population into a social circle of $N-m$ individuals and a population of $m$ stereotyped individuals, who are in turn divided into into $G$ sub-groups, each comprising $n=m/G$ individuals. A player keeps track of the identity of others and the amount of help they have received from them. And so a focal player's strategy requires them to recall the identity and behavior of members each of the $N-m$ members of their social circle, along with the sub-group assignment of each of the $m$ other players in the population and the number of times they have been helped by that subgroup. 

The total information required to store the identity and behavior of each individual in the population is

\[
I=N\log_2[A^*(N-m+1)(G+1)]+G\log_2(n+1)+(N-m)\log_2 2
\]
\\
where $A^*$ is the information required to remember the identity of each individual in the population, so that the first term captures the information required to remember the membership of all individuals to with respect to $N-m$ groups of 1 within their social circle and to $G$ stereotype groups of $n$ individuals. The second term is the information required to remember the amount of help the focal player received from each stereotype group, while the final term is the information required to remember the help received from members of the social circle. We make the simplifying assumption that the cost per individual identity remembered, $A^*$, is constant and so does not impact the evolutionary dynamics. The total information $I$ can be broken down into the component due to social circle members and the component due to stereotyped individuals. This is the form given in main text Eq. 3 where we express the information per individual, i.e. we divide through by $N$ to make the adaptive dynamic limit tractable. Similarly we write $A=A^*N^2$ in main text Eq. 3 and assume terms $1/N\ll 1$. As described in the main text, we assume that the need to remember this information comes at a cognitive cost \cite{Milinski:1998aa,Stevens05}, such that fitness decreases exponentially with the information required for a strategy. We assume that the size of the cognitive cost $C_m$ is constant across context i.e. it is the same for remembering stereotype groups and social circle members.

\subsection*{Evolutionary singular social circles}

We can now compute the selection gradient from main text Eq. 4

\begin{eqnarray}
   \nonumber \frac{\partial w_i}{\partial x_m^i}\Bigg |_{x_m^i=x_m}=-(B-C)(1-c_m)^{I^*_{c}(x_m)}
    +(B-C)(1-c_m)^{I^*_{s}(x_m,g)}\Pi_c(x_m,g)+\\
   \nonumber (1-x_m)(B-C)(1-c_m)^{I^*_{c}(x_m)}\ln(1-c_m)\frac{\partial I^*_c}{\partial x_m^i}\Bigg |_{x_m^i=x_m}+\\
    x_m(B-C)(1-c_m)^{I^*_{s}(x_m,g)}\Pi_c(x_m,g)\ln(1-c_m)\frac{\partial I^*_s}{\partial x_m^i}\Bigg |_{x_m^i=x_m}
\end{eqnarray}
\\
where we have set $c_m=1-(1-C_m)^{1/\ln(2)}$ with $I^*_c=\ln(2)I_c$ and $I^*_s=\ln(2)I_s$ so that

\begin{equation}
    \frac{\partial I^*_c}{\partial x_m^i}\Bigg |_{x_m^i=x_m}=-1-\frac{1}{1-x_m}
\end{equation}
\\
and

\begin{equation}
    \frac{\partial I^*_s}{\partial x_m^i}\Bigg |_{x_m^i=x_m}=\frac{g}{g+x_m}
\end{equation}
\\
and the condition for equilibrium (i.e zero selection gradient) is

\begin{eqnarray}
 \Pi_c(x_m) =
(g+x_m)(1 - c_m)^{I^*_c(x_m)-I^*_s(x_m)} \frac{1+\ln (1- c_m) + (1 - x_m) \ln(1 - c_m)}{(g + x_m) + g x_m \ln(1 - c_m)}
 \end{eqnarray}
 \\
 
 where we can see from Eq 17-18 that $(1 - c_m)^{I^*_c(x_m)-I^*_s(x_m)}$ is monotonically increasing in $x_m$, while the term $\frac{1+\ln (1- c_m) + (1 - x_m) \ln(1 - c_m)}{(g + x_m) + g x_m \ln(1 - c_m)}$ is also monotonically increasing in $x_m$. Since $\Pi_c(x_m)$ is monotonically decreasing in $x_m$ (see Methods), Eq. 19 has just one solution in $x_m$, indicating that there is a unique singular social circle size for the adaptive dynamics, $\bar{x}_m$.  The precise value of $\bar{x}_m$ must be determined numerically in general. Its variation with the system parameters is shown in Figure S4.
 
  In the limit $C_m\to 1$, Eq. 19 has solution $\Pi_c\to0$, which according to main text Eq. 2 and Eq. 30 occurs when $n\to\infty$, which implies $x_m\to 1$ for fixed $g$, i.e. selection acts to maximize the number of people stereotyped (see mean text Figure 4).  
 
 Setting $x_m=1$ in Eq. 16 we recover

 \begin{eqnarray}
   \nonumber \frac{\partial w_i}{\partial x_m^i}\Bigg |_{x_m^i=x_m}=-(B-C)(1-c_m)^{I^*_{c}(x_m)}\left(1+\ln(1-c_m)\right)+\\
\nonumber    +(B-C)(1-c_m)^{I^*_{s}(x_m,g)}\Pi_c(x_m,g)\left(1+\ln(1-c_m)\frac{g}{g+1}\right)
\end{eqnarray}
\\
 Since $I^*_c\to-\infty$ as $x_m\to 1$ it is the first term that dominates, and the selection gradient becomes large and positive, i.e. there is no singular strategy when $c_m\to 1$, and selection instead acts to maximize $x_m$.
 
 \begin{figure}
\centering
\includegraphics[scale=0.4]{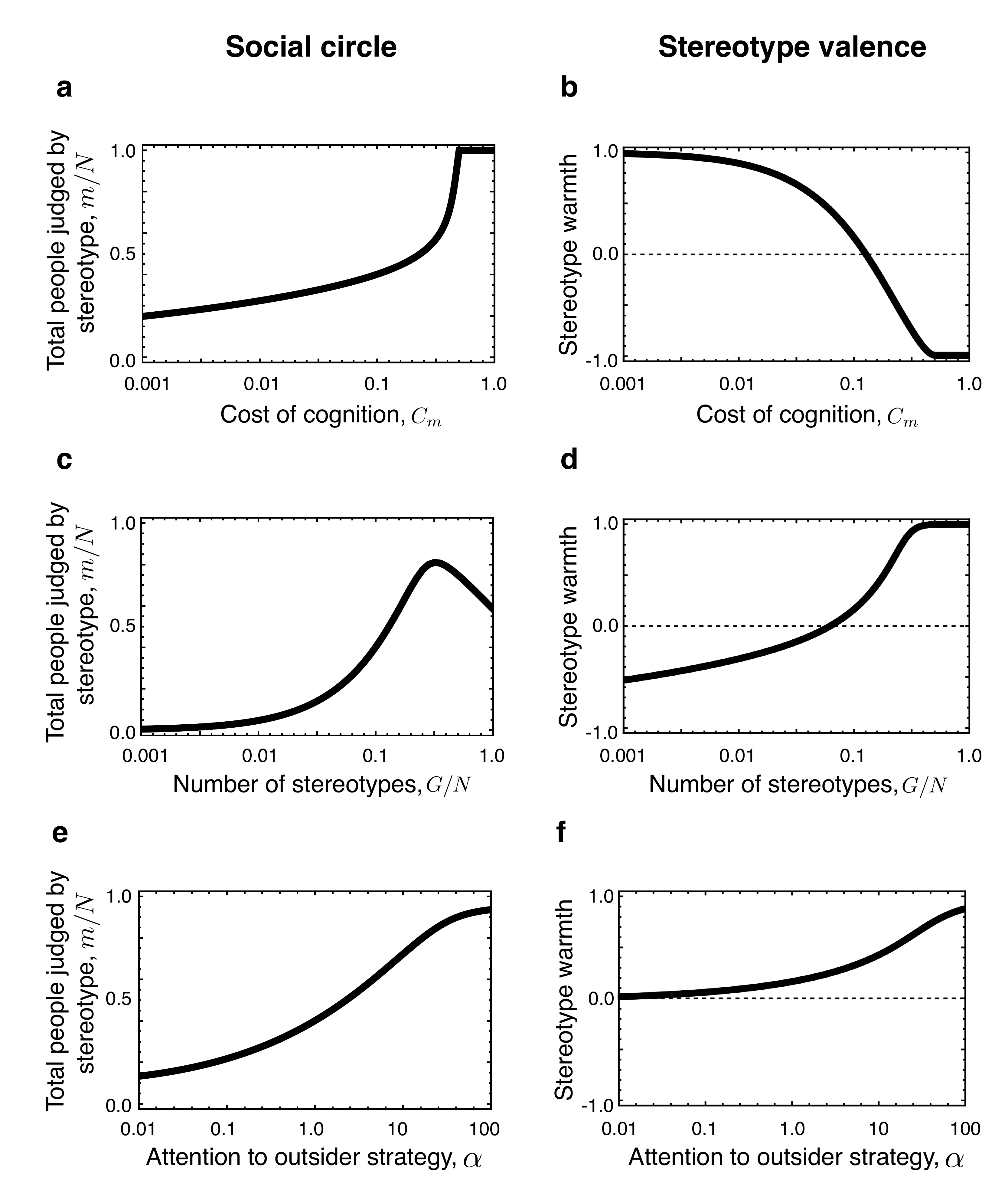}
\caption*{Figure S4: \textbf{Evolutionary singular social circle.} We calculated the evolutionary singular strategy $\bar{x}_m=m/N$ (left column) and the associated stereotype warmth (right column) as a function of (a-b) cognitive costs $C_m$, (c-d) number of stereotype groups $G/N$ and (e-f) attention to out-group strategy, $\alpha$. Singular strategies are calculated numerically for the adaptive dynamics as described in the main text and supplement. Here $B=5$, $C=1$ and $\beta=0.0011$ along with $C_m=10^{-2}$, $G/N=0.1$ and $\alpha=0.5$ unless otherwise stated.}
\end{figure}
 
 \subsection*{Uninvadability of the singular social circle}
 
When $c_m<1$ we can assess the stability of the unique equilibrium by looking at the second derivative of the selection gradient

 \begin{eqnarray}
   \nonumber \frac{\partial^2 w_i}{\partial (x_m^i)^2}\Bigg |_{x_m^i=\bar{x}_m}=(B - C)\ln(1 - c_m) \Bigg[(1 - c_m)^{I^*_c}\frac{ 3 - 2\bar{x}_m + (2 - \bar{x}_m)^2 \ln(1 - c_m)}{1-\bar{x}_m} + \\
   (1 - c_m)^{I^*_s}\frac{g  (2 g + \bar{x}_m + g \bar{x}_m \ln(1 -c_m))}{(g + \bar{x}_m)^2}\Pi_c\Bigg]
\end{eqnarray}
\\
If we Taylor expand Eq. 20 in $c_m$ we recover

 \begin{eqnarray}
    \frac{\partial^2 w_i}{\partial (x_m^i)^2}\Bigg |_{x_m^i=\bar{x}_m}=-(B - C) c_m \Bigg[\frac{ 3 - 2\bar{x}_m }{1-\bar{x}_m} + 
   \frac{g  (2 g + \bar{x}_m )}{(g + \bar{x}_m)^2}\Pi_c\Bigg]+O(c_m^2)
\end{eqnarray}
\\
and so the singular strategy $\bar{x}_m$ cannot be invaded when $c_m\ll 1$. When $c_m\to 1$ the selection gradient remains positive and $x_m = 1$ as shown above, meaning that we do need to compute the second derivative to assess stability.

  \subsection*{Convergence stability of the singular social circle}
 
Finally we explore the convergence stability of the evolutionary singular strategy $\bar{x}_m$.

\begin{eqnarray}
   \frac{\partial}{\partial x_m}\left(\frac{\partial w_i}{\partial x_m^i}\Bigg |_{x_m^i=x_m}\right)=
    (B-C)(1-c_m)^{I^*_{s}(x_m,g)}\frac{\partial \Pi_c}{\partial x_m}\left(1+x_m\ln(1-c_m)\frac{gx_m}{g+x_m}\right) + \frac{\partial^2 w_i}{\partial (x_m^i)^2}\Bigg |_{x_m^i=\bar{x}_m}
\end{eqnarray}
\\
The first term is only positive if

\begin{equation}
    1+\ln(1-c_m)\frac{gx_m}{g+x_m}<0
\end{equation}
 \\
 which requires $c_m>1-e^{-2}$. And so the system is also convergent stable when $c_m\ll1$.

 \subsection*{Optimal group size}
 
 The adaptive dynamics analysis performed for social circle size, $\bar{x}_m$, is not appropriate for the number of groups $g$, since groupings are assumed to be mutually shared by all members of the population. It is nonetheless interesting to consider the effect of the number of groupings on the equilibrium fitness of the system, evaluated at the evolutionary singular strategy $\bar{x}_m$. We explored this numerically and characterized the optimal group size $\bar{g}\leq \bar{x}_m$ (main text Figure 3) and visualized the selection gradient for group size and social circle size (Figure S5).
 
  \begin{figure}
\centering
\includegraphics[scale=0.4]{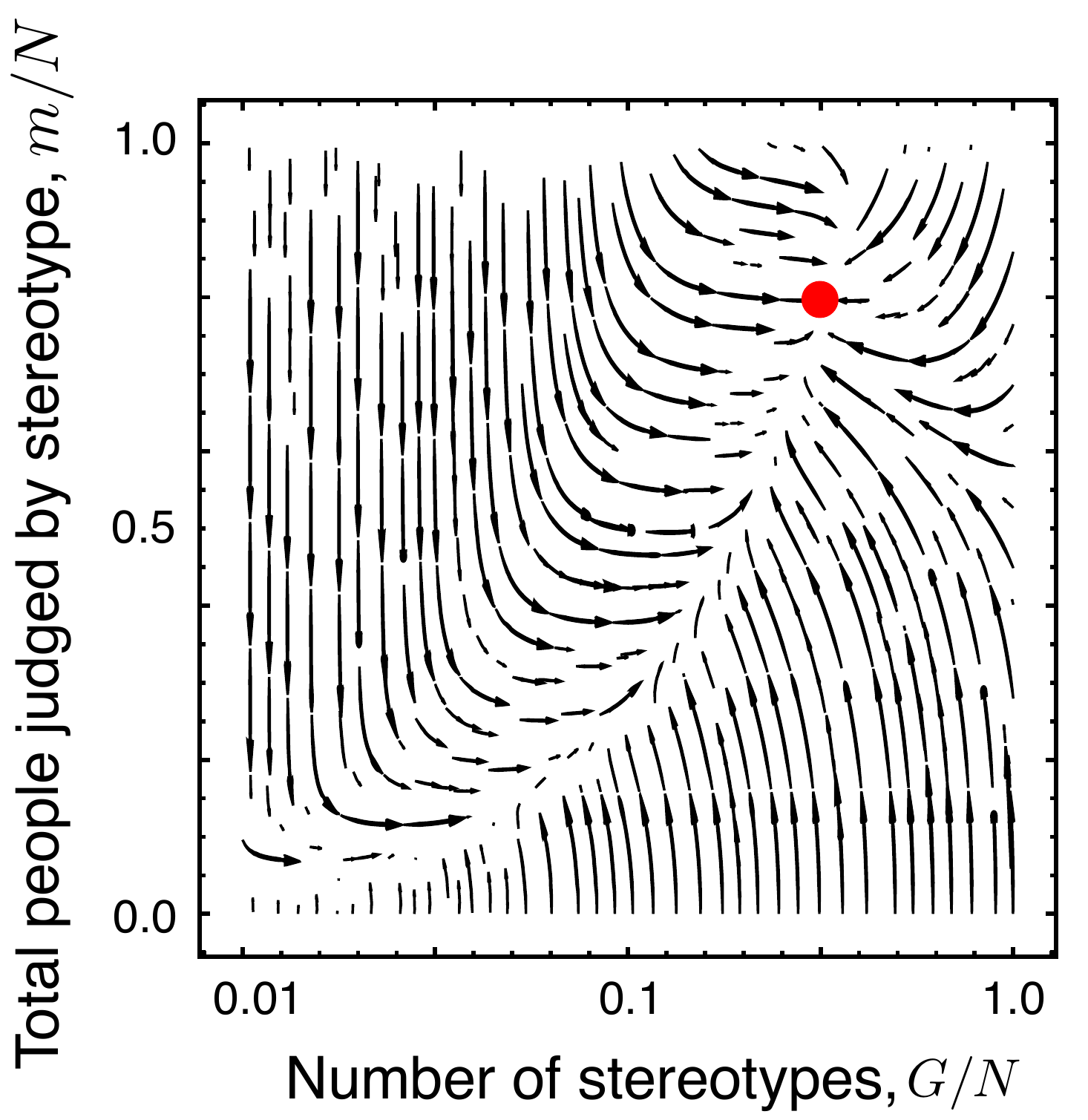}
\caption*{Figure S5: \textbf{Optimal group size.} We calculate the direction of the selection gradient acting on the social circle size $m/N$ and on the group size $G/N$ under the adaptive dynamics model. Arrows show the direction of selection, while the red dot shows the location of the unique equilibrium for the system, for $B=5$, $C=1$, $C_m=10^{-2}$ with $\beta=0.0011$ and $\alpha=0.5$ }
\end{figure}

\subsection*{Imitation of out-groups}

Next we explore the effect of varying the rate of out-group imitation, $\alpha$, on the evolutionary dynamics of stereotyping.  Setting

\[
\gamma=\frac{n-1}{n-1+\alpha G}
\]
\\
as the relative rate of copying members of the same stereotype group, we see from main text Eq. 30 that when $\alpha=0$, so that $\gamma=1$ and players always copy within stereotype group behavior, we recover

\begin{equation}
  \rho=1-  \frac{ 
  \sqrt{4 C^2 n^2}}{2Cn}=0
\end{equation}
\\
i.e. no robust cooperative strategies exist in this limit, and stereotypes are always negative. More generally, the warmth of stereotypes decreases as out-group imitation decreases, as shown in Figure S4f.

\subsection*{Interactions within social circles}

Next we consider the effect of within social circle interactions becoming less than perfectly cooperative. In general, direct reciprocity produces less than perfect cooperation in the long run when $B<2$ in the donation game \cite{Stewart:2014aa}, but will produce high levels of cooperation otherwise. Nonetheless it could in general be the case that the benefits of cooperation differ between social circle members and stereotype groups (for example, within a social circle there may be costly cooperation enforcement mechanisms that reduce the net benefit of cooperation). To explore this possibility, we vary the average within social circle benefit received $B^*<B$, and explore the effects of varying $B^*$ on the adaptive dynamics of stereotypes. Relaxing this assumption does not qualitatively impact our analysis of evolutionary singular strategies since it can be accounted for by dividing $\Pi_c$ in Eq. 16 by a constant factor $(B-C)/(B^*-C)$.

The effect on the quantitative value of the singular strategy of varying $B^*$ is shown in Figure S6 where we observe that increasing $B^*>B$ quickly causes stereotyping to vanish, as we might expect. However decreasing $B^*<B$ causes the proportion of stereotyped individuals $x_m$ to quickly increase to 1, while the number of stereotype groups $g$ initially increases, before decreasing, indicating increasingly coarse stereotypes as within social circle interactions become less beneficial.

  \begin{figure}
\centering
\includegraphics[scale=0.5]{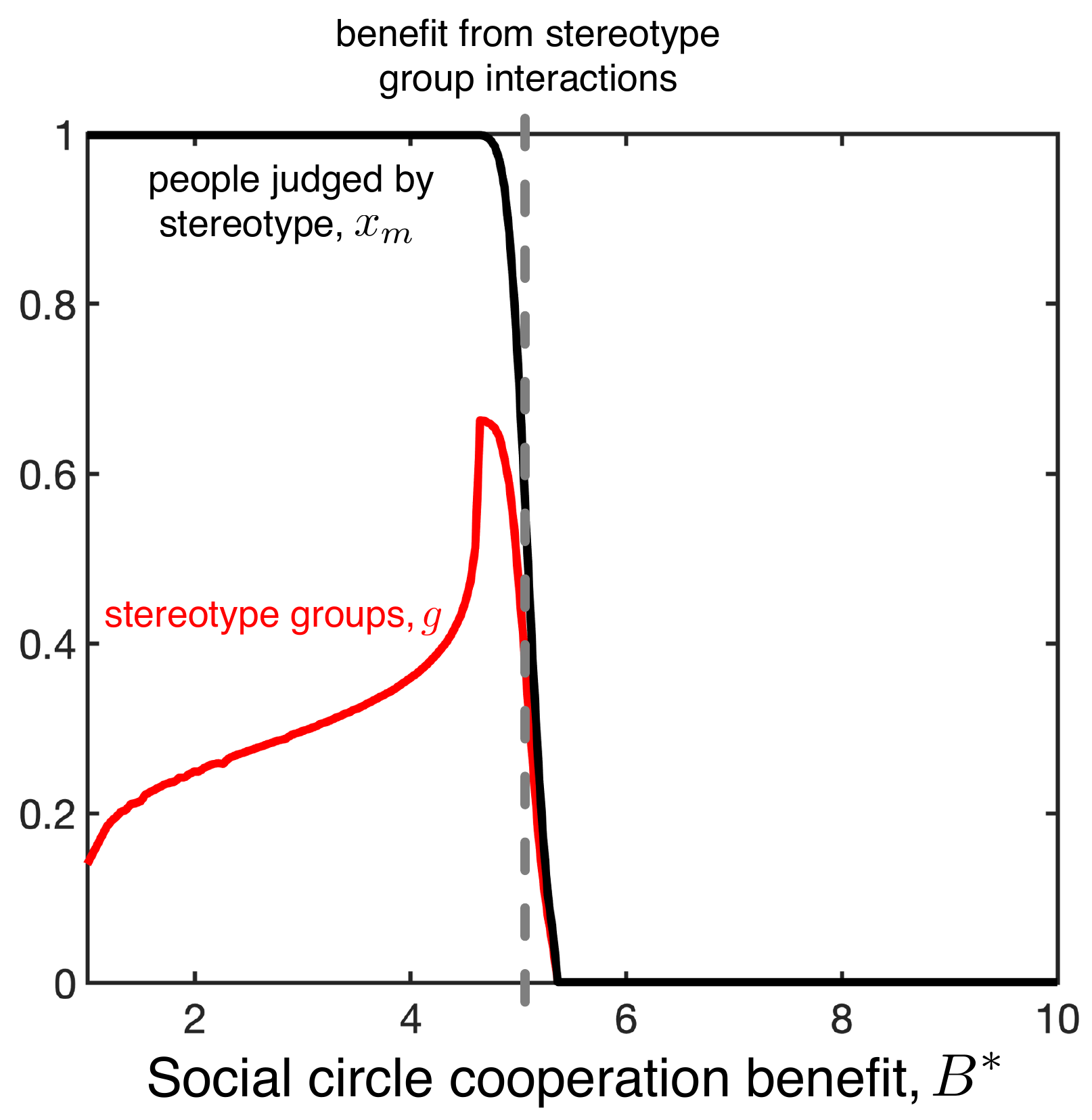}
\caption*{Figure S6: \textbf{Benefit of social circle interactions.} We calculated the evolutionary stable social circle size, $1-x_m$ as a function of the number of stereotypes, $g$, where we have set $x_m=\frac{m}{N}$ and $g=\frac{G}{N}$, and taken the limit $N\to\infty$. We then calculated the value of $g$ that maximizes fitness, to give the evolutionary optimal social circle size and number of stereotype groups for a given set of parameters. Evolutionary optimal social circle size (black line) and number of stereotype groups (red line) as a function of $B^*$. Evolutionary optima are calculated numerically with $B=5$ (vertical grey line) and $C=1$, $\alpha=0.5$ and $\beta=0.0011$}
\end{figure}

\section*{Stereotypes based on common-knowledge}

Finally we relax the assumption that the average behavior of a group, used to determine a focal player's strategy as described in Eq. 1, depends only on the experience of that focal player in interacting with the group and instead assume that the player averages the experience of all members of their stereotype group, i.e.

\begin{equation}
p_k=s\sum_{i=1}^n\frac{k_i}{n^2}+r
\end{equation}
\\
where $i$ indexes each member of the focal individual's stereotype group. This change does not change the invasion conditions given in main text Eqs. 24-30, since in the infinitely repeated game all members of a stereotype group experience the same average rate of cooperation from a given out-group $j$. However we can also consider the case where common knowledge occurs at the population level, i.e.

\begin{equation}
p_k=s\sum_{i\neq j}\frac{k_i}{n(N-n)}+r
\end{equation}
\\
where $i$ indexes all members of the population not belonging to to the out-group $j$ being considered. Under this model, in a large population $N\gg1$, a resident cooperative strategy invaded by a non-cooperative strategy will cause the resident strategy in the group containing the mutant to continue cooperating with out-groups even as those out-groups withdraw cooperation, i.e $\left<j_r\right>=n$ in main text Eqs. 13-14. Solving main text Eq. 13 and Eq. 14 for $\left<j_g\right>$ and $\left<j_m\right>$ gives

\begin{equation}
    \left<j_m\right>=\frac{n(s_m+r_m)-s_ms_r}{n-s_r s_m }
\end{equation}
\\
and

\begin{equation}
    \left<j_g\right>=\frac{n-s_r(1-r_m)}{n- s_r s_m }
\end{equation}
\\
and solving to find the robustness $\rho$ gives

\begin{equation}
  \rho=1-\frac{n}{1-\gamma}\frac{C}{B}
\end{equation}
\\
We can now compare Eq. 29 to main text Eq. 30, i.e. the case with and without common knowledge of stereotypes. This is shown in Figure S7 where we see the robustness of cooperative strategies with population-level common knowledge is markedly less than the robustness of stereotypes based on individual or group-level knowledge.

  \begin{figure}
\centering
\includegraphics[scale=0.4]{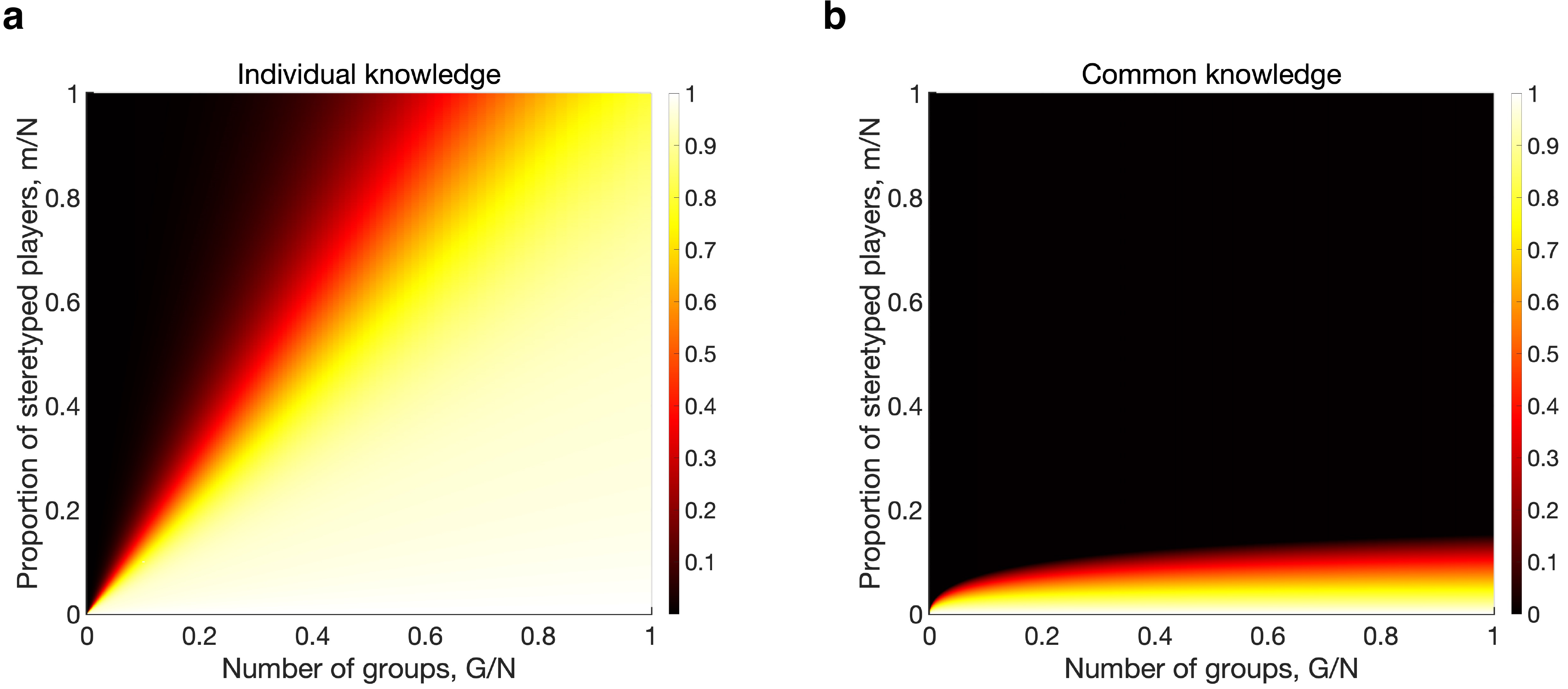}
\caption*{Figure S7: \textbf{Individual vs common knowledge.} We calculated the robustness $\rho$ (colors) as a function of $m/N$ and $G/N$  for a) stereotypes formed based on individual knowledge and b) for stereotypes based on population level common knowledge. Here $B=5$, $C=1$ and $\beta=0.0011$ along with $C_m=10^{-2}$, $G/N=0.1$ and $\alpha=0.5$ unless otherwise stated.}
\end{figure}

\section*{Individual based simulations}
 C++ and Matlab code used to generate simulation and numerical results is available via github. The constant $\beta$ was calculated numerically based on the value of $n$ that produced $\rho=0.5$ given $B$ and $C$.


\end{document}